\documentclass[journal]{vgtc}                     % final (journal style)
\onlineid{0}

\vgtcpapertype{Analytics \& Decision}

\newcommand{\revised}[1]{{\color{black} #1}}

\usepackage{tabu}                      % only used for the table example
\usepackage{booktabs}                  % only used for the table example
\usepackage{lipsum}                    % used to generate placeholder text
\usepackage{mwe}                       % used to generate placeholder figures
\usepackage{multirow} % For multi-row cells

\usepackage{mathptmx}                  % use matching math font
\usepackage{setspace}
\newcommand{\schemaname}[1]{{\fontfamily{txtt}\selectfont #1}}
\usepackage{xspace, xpunctuate}
\usepackage{listings}  % package for code highlighting
\usepackage{enumitem}  % compress the space in the enumeration environment

\usepackage{algorithm}
\usepackage{algpseudocode}

\lstdefinelanguage{json}{
    basicstyle=\fontfamily{txtt}\selectfont\footnotesize,
    numbers=left,
    numberstyle=\color{gray}\footnotesize\ttfamily,
    numbersep=4pt,
    tabsize=2,
    showstringspaces=false,
    breaklines=true,
    xleftmargin=10pt,
    literate=
     *{0}{{{\color{numb}0}}}{1}
      {1}{{{\color{numb}1}}}{1}
      {2}{{{\color{numb}2}}}{1}
      {3}{{{\color{numb}3}}}{1}
      {4}{{{\color{numb}4}}}{1}
      {5}{{{\color{numb}5}}}{1}
      {6}{{{\color{numb}6}}}{1}
      {7}{{{\color{numb}7}}}{1}
      {8}{{{\color{numb}8}}}{1}
      {9}{{{\color{numb}9}}}{1}
      {:}{{{\color{punct}{:}}}}{1}
      {,}{{{\color{punct}{,}}}}{1}
      {\{}{{{\color{delim}{\{}}}}{1}
      {\}}{{{\color{delim}{\}}}}}{1}
      {[}{{{\color{delim}{[}}}}{1}
      {]}{{{\color{delim}{]}}}}{1},
}

\lstdefinestyle{mystyle}{
  basicstyle=\footnotesize\ttfamily,
    numbers=none,
  frame=none,
  columns=flexible,
  xleftmargin=0pt,
  aboveskip=-1pt,
  belowskip=0pt
}

\title{ProactiveVA: Proactive Visual Analytics with LLM-Based UI Agent}
\author{%
  \authororcid{Yuheng Zhao}{0000-0003-1573-8772},
  \authororcid{Xueli Shu}{0009-0003-5058-0541},
  \authororcid{Liwen Fan}{0009-0007-3568-1271},
  \authororcid{Lin Gao}{0009-0004-1613-1774},
  \authororcid{Yu Zhang}{0000-0002-9035-0463},
  \authororcid{Siming Chen}{0000-0002-2690-3588}
}

\authorfooter{
  \item
  	Y. Zhao, X. Shu, L. Fan, L. Gao, and S. Chen are with Fudan University. S. Chen is also with Ji Hua Laboratory and Shanghai Key Laboratory of Data Science. S. Chen is the corresponding author (simingchen@fudan.edu.cn)
  \item
  	Yu Zhang is with the University of Oxford.
}

\abstract{%
Visual analytics (VA) is typically applied to complex data, thus requiring complex tools.
While visual analytics empowers analysts in data analysis, analysts may get lost in the complexity occasionally. 
This highlights the need for intelligent assistance mechanisms.
However, even the latest LLM-assisted VA systems only provide help when explicitly requested by the user, making them insufficiently intelligent to offer suggestions when analysts need them the most.
We propose a ProactiveVA framework in which LLM-powered UI agent monitors user interactions and delivers context-aware assistance proactively.
To design effective proactive assistance, we first conducted a formative study analyzing help-seeking behaviors in user interaction logs, identifying when users need proactive help, what assistance they require, and how the agent should intervene.
Based on this analysis, we distilled key design requirements in terms of intent recognition, solution generation, interpretability and controllability. Guided by these requirements, we develop a three-stage UI agent pipeline including perception, reasoning, and acting. 
The agent autonomously perceives users' needs from VA interaction logs, providing tailored suggestions and intuitive guidance through interactive exploration of the system. We implemented the framework in two representative types of VA systems, demonstrating its generalizability, and evaluated the effectiveness through an algorithm evaluation, case and expert study and a user study. We also discuss current design trade-offs of proactive VA and areas for further exploration.

}


\graphicspath{{images/}{./}}

\begin{document}

%%%%%%%%%%%%%%%%%%%%%%%%%%%%%%%%%%%%%%%%%%%%%%%%%%%%%%%%%%%%%%%%
%%%%%%%%%%%%%%%%%%%%%% START OF THE PAPER %%%%%%%%%%%%%%%%%%%%%%
%%%%%%%%%%%%%%%%%%%%%%%%%%%%%%%%%%%%%%%%%%%%%%%%%%%%%%%%%%%%%%%%


\maketitle
\begin{spacing}{0.971}
\section{Introduction}
\label{sec:introduction}

\begin{figure}[h!]
    \centering
    \includegraphics[width=\linewidth]{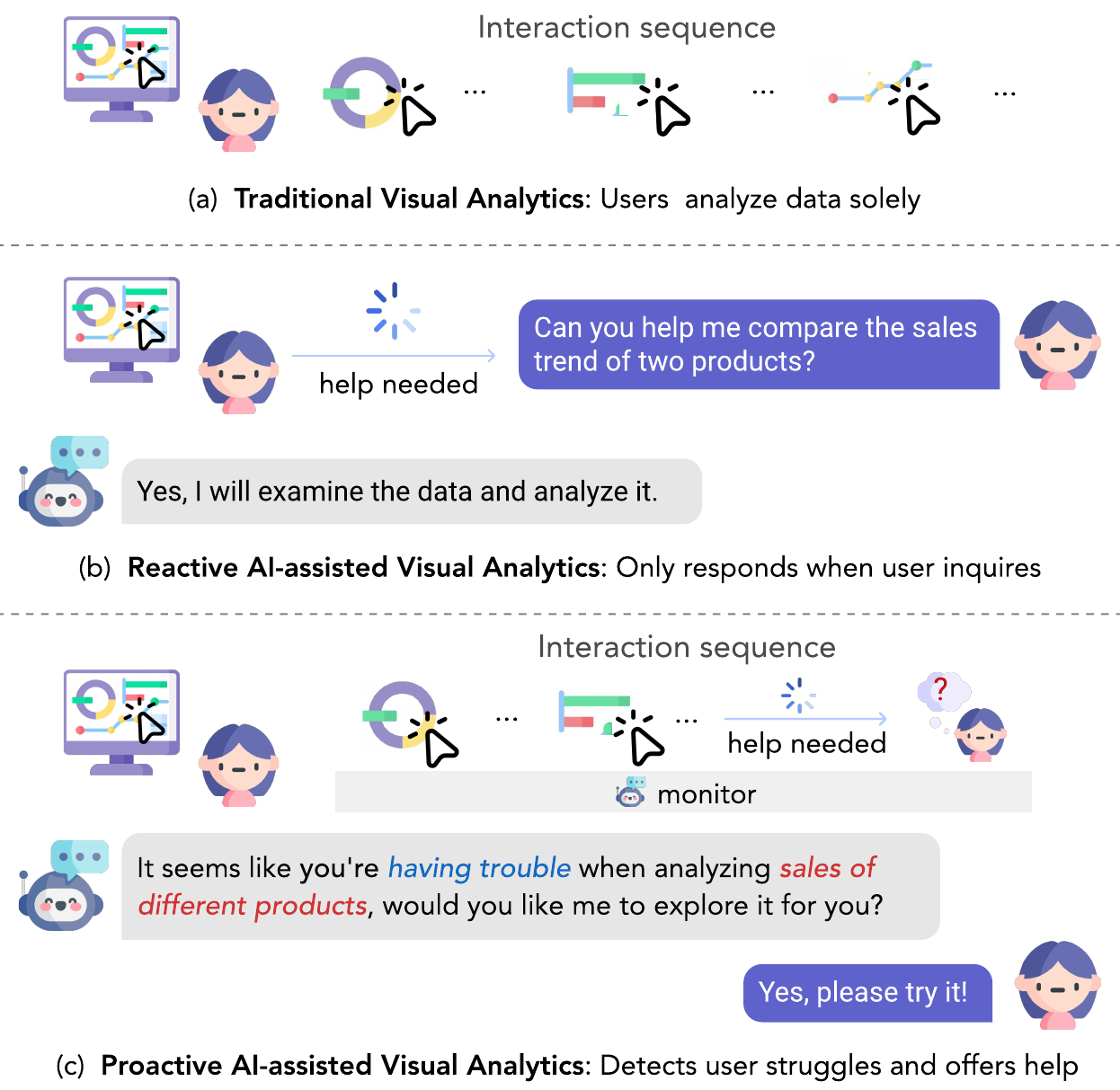}
    \caption{Illustration of three different visual analytics paradigms.}
    \label{fig:comparison}
    \vspace{-20pt}
\end{figure}

% 说明 VA 的终极愿景是实现人机协同
The ultimate vision and goal of visual analytics (VA) is to enable intelligent human-AI collaboration \revised{in analyzing complex data}~\cite{thomas2005illuminating, KeimVisual2008c}. 
% 介绍传统 VA 模式，强调用户主导交互，以及面临的挑战
\revised{In the traditional VA paradigm (Fig.~\ref{fig:comparison}a), users interact with algorithms and visualizations to make sense of data. However, when dealing with complex systems, they may occasionally become overwhelmed by the system’s complexity}~\cite{ceneda2016characterizing}.
% 引出响应式智能助手作为缓解复杂性的早期手段，强调其依赖显式请求
To facilitate the analysis process, reactive artificial intelligence (AI) assistants are developed to provide on-demand service but rely on explicit user requests~\cite{ZhaoLEVA2024, KimPhenoFlow2024c} (Fig.~\ref{fig:comparison}b).

% 引入 mixed-initiative 系统的目标：从响应式走向预测式协作
Mixed-initiative systems further aim to enable AI assistants to proactively predict user needs and provide context-sensitive support for collaboration~\cite{HorvitzPrinciples1999a}.
% 承接上一句，说明已有多种方法被提出用于 mixed-initiative VA
Various approaches have been proposed to realize this vision for VA over the last years.
% 举例第一类方法：基于规则的系统，说明其原理和优点（可主动推荐）
\revised{On the one hand, rule-based systems~\cite{CookMixedinitiative2015, SperrleLearning2021, HeerInteractive2012}, such as DataSite~\cite{CuiDataSite2018a}, can actively recommend insights, based on predefined proactive specifications.
% 点出 rule-based 系统的局限：泛化性差，适应性不足
However, the reliance on manually defined rules and intervention points limits the generalizability and adaptability to diverse analytical contexts.
% 举例第二类方法：学习型系统，从交互历史中学习行为预测
On the other hand,} learning-based systems can predict next-step actions, learning from historical interaction sequences~\cite{LiDiverse2022, wall2019markov, DabekGrammarbased2017}.
% 点出 learning-based 方法的两个问题：时机掌控差、可解释性差
\revised{While these methods capture users' implicit intention, they often deliver constant recommendations without regard to timing and operate as black-box recommenders, which may interrupt users’ reasoning processes and undermine trust in the system.
% 总结两类方法局限，引出下一步研究目标
These limitations highlight the need for proactive agents that can autonomously determine when and what to help, while maintaining interpretability and controlability within dynamic analytic workflows.}

% 概括提出的方案：一个主动式框架 + UI agent + 关键能力
To address these limitations, we propose a ProactiveVA framework that uses an autonomous UI agent to monitor user behaviour, assess the difficulties users face, and provide proactive and explainable assistance at the right moment (Fig.~\ref{fig:comparison}c). 
% 指出研究挑战：用户行为和思维难以预测
\revised{Designing such a system is challenging because users’ thinking processes and actions are complex and difficult to predict.} 
% 提出研究问题框架：三大问题 RQ1-3
We decompose this challenge into three core research questions: \textit{When do users need help (RQ1)? What kinds of help do they need (RQ2)? How should the agent intervene (RQ3)?}
% 引出方法：开展形成性研究，说明采集方式
Therefore, we first conducted a formative study, observing user interactions, collecting interaction logs, think-aloud protocols, and interview data.
% 调研结果（RQ1+RQ2），并指出其与已有研究的一致性
After analyzing the collected data and reviewing literature, we derive that the timing and assistance strategy, including three aspects: onboarding tips, exploration guidance, and error reminders, is aligned with a previous study~\cite{ZhaoLEVA2024}.
% 调研结果（RQ3）: 用户偏好的介入方式
\revised{Moreover, users would prefer assistance that is timely yet non-intrusive, visually grounded, and aligns with their current workflow.}
% 引出系统设计目标，呼应三个研究问题
Based on these findings, we designed a proactive UI agent to autonomously provide timely help for users to facilitate mixed-initiative VA scenarios. 
% 具体说明系统如何感知何时介入（RQ1）
Specifically, to perceive when users need help, \revised{the UI agent monitors user interactions to detect confusion events by analyzing temporal features (e.g., extended thinking time), task phases (e.g., onboarding vs. mid-analysis), and user behavior context (e.g., interaction data).}
% 说明系统如何理解用户意图并选择正确帮助（RQ2）
To provide the appropriate help, we infer the user's intent by analyzing interaction patterns and context, allowing the agent to suggest relevant actions or guidance. 
% 说明系统如何以可理解方式呈现帮助（RQ3）
Finally, to align the agent's recommendations with the user's cognitive process, we provide step-by-step guidance with a combination of text and visuals in a subtle, accessible format that integrates seamlessly into the user's workflow.

Our contributions are as follows:

\begin{itemize}[leftmargin=*]
    \item We identify key requirements for \revised{proactive assistance in visual analytics} through a formative study, revealing requirements for context-aware, non-disruptive assistance.
    \item We propose a proactive visual analytics framework supported by an LLM-based UI agent. The agent autonomously determines when and how to provide proactive assistance.
    \item We implement the framework in two types of visual analytics systems, demonstrating its generalizability. We evaluate the effectiveness through an algorithm evaluation, case and expert study, and a user study. We also discuss the current design trade-offs of proactive VA and areas for further exploration.
\end{itemize}

\section{Related Work}
\label{sec:related-work}

In this section, we reviewed previous studies related to proactive agents for VA, including mixed-initiative VA systems, user behaviour understanding, guidance for VA and LLM-based proactive agents.

\subsection{Mixed-Initiative Visual Analytics}

\revised{Mixed-initiative system is one where both the human and the agent can take the initiative to contribute to the task, and control may shift dynamically during the interaction~\cite{HorvitzPrinciples1999a}.
In VA, the aim is to support analytical workflows through human–agent collaboration, assisting in visualization understanding, data exploration, insight generation~\cite{thomas2005illuminating, KeimVisual2008c, PisterIntegrating2021a}.}
A key enabler of such collaboration is to understand user intent during data exploration~\cite{XuSurvey2020}. 
Prior work has explored this from various levels: identifying behavioral patterns such as scanning, flipping, or comparing~\cite{GotzCharacterizing2009, GotzBehaviordriven2009}, segmenting actions into meaningful sequences~\cite{AndrienkoSeeking2022}, and classifying tasks at different levels of abstraction~\cite{BrehmerMultiLevel2013, BoukhelifaCase2021, GuoCase2016}. These methods form the basis for adaptive assistance.

Building on behavioral understanding, various systems have proposed proactive or mixed-initiative guidance mechanisms. One line of work uses machine learning to predict user intent or next steps~\cite{BrownFinding2014, DabekGrammarbased2017}. For instance, Li et al.~\cite{LiDiverse2022} apply LSTM models to model interaction histories and recommend visual transitions. Another line of work employs rule-based or task-driven strategies. Cook et al.~\cite{CookMixedinitiative2015} and Guo et al.~\cite{GuoProactive2011b} use pre-defined rules to recommend data transformations. Cui et al.~\cite{CuiDataSite2018a} proactively compute insights and push recommendations through a feed, while Sperrle et al.~\cite{SperrleLearning2021} learn contextual rules from user feedback to drive agent suggestions.
\revised{Miksch et al.~\cite{perez2023guided} design a guided system for spatio-temporal image selection, where the system suggests relevant story elements during exploration.}
\revised{While prior systems have made progress in proactive visual analytics, they often lack awareness of when users need help and the ability to reason about tasks to generate context-appropriate strategies. Most rely on fixed triggers and are tailored to specific domains, which limits their generalizability.
To address these challenges, we introduce an LLM-powered proactive UI agent that integrates behavior-aware need detection, contextual reasoning, and autonomous interaction to provide assistance.}

\subsection{LLM Agent for Visual Analytics}

\revised{The emergence of large language models (LLMs) has sparked new directions in visual analytics, particularly in automating reasoning and content generation.} Early efforts focus on enabling LLMs to generate visualizations and analysis content based on user input. For example, ChartGPT~\cite{TianChartGPT2023b} and LIDA~\cite{DibiaLIDA2023} use LLMs to convert natural language prompts into visual representations. InsightLens~\cite{weng2025insightlens} and AVA~\cite{liu2024ava} build agent-based frameworks to support data analysis and visualization fine-tuning, respectively. LightVA~\cite{ZhaoLightVA2024} \revised{further integrates task decomposition and data mining, building a lightweight generative framework.}
\revised{Other systems aim to recommend insights and annotations in VA system based on user queries.} LEVA~\cite{ZhaoLEVA2024} observes user exploration patterns and suggests potentially relevant tasks using an insight function library. Gao et al.~\cite{GaoFineTuned2024} propose a fine-tuned LLM framework that recommends questions to support self- learning in an educational VA system.

\revised{Moving toward proactive assistance, recent systems explore how to anticipate user needs during analysis.} ArticulatePro~\cite{TabalbaArticulatePro2024} continuously listens to spoken conversations and generates relevant visualizations on the fly. \revised{While these systems offer flexible input and proactive generation, they do not explicitly consider the timing of assistance or the evolving cognitive state of the user.}
\revised{To address these gaps, we propose a proactive agent that can autonomously perceive when users may need help, reason about what to suggest, and act through the interface in real time.}

\subsection{Proactive LLM Agent}

\revised{Proactivity is commonly defined along three dimensions: anticipation of user needs, initiation of actions without explicit commands, and targeting assistance toward the user’s current or future goals~\cite{grant2008dynamics, peng2019design}.}
Most existing proactive LLM agents are developed in dialogue settings~\cite{DengSurvey2023a}. \revised{These agents anticipate user intent~\cite{LiuProactive2024}, provide gentle error correction~\cite{ZarghamUnderstanding2022a}, or proactively seek relevant information~\cite{zhao2024generating}. For example, ComPeer~\cite{LiuComPeer2024} reflects on important conversational events to plan proactive care, while Liu et al.~\cite{LiuProactive2024} enable agents to internally formulate intentions and decide when to contribute.}

With improved tool-use capabilities, \revised{ LLM agents now also initiate actions across domains} such as programming~\cite{ross2023programmer}, writing~\cite{kadoma2024role}, travel planning~\cite{ZhangAskbeforePlan2024}, and smart homes~\cite{OhBetter2024a}. For instance, Chen et al.~\cite{ChenNeed2024} develop a proactive coding assistant that offers suggestions based on real-time code context. Lu et al.~\cite{LuProactive2024} enhance LLM agents by training on desktop activity to infer when help is needed.
\revised{However, most existing agents target domains with constrained user goals and linear workflows. In contrast, VA presents unique challenges across all three dimensions of proactivity: analytical tasks are dynamic, user behaviors are diverse and multimodal, and the timing of support is highly sensitive to context.}
\revised{To address these challenges, we build on the concept of UI agents, which are capable of perceiving and interacting with user interfaces~\cite{zhang2024ufo, ZhouWebArena2024, LaiAutoWebGLM2024}. Our approach enables three key aspects of proactivity: anticipation of user needs through behavior-aware detection, autonomous initiation of actions through LLM-driven planning, and targeted assistance aligned with users’ evolving analytical goals.}
% \section{Design Requirements}
\begin{figure*}[h!]
    \centering
    \includegraphics[width=\linewidth]{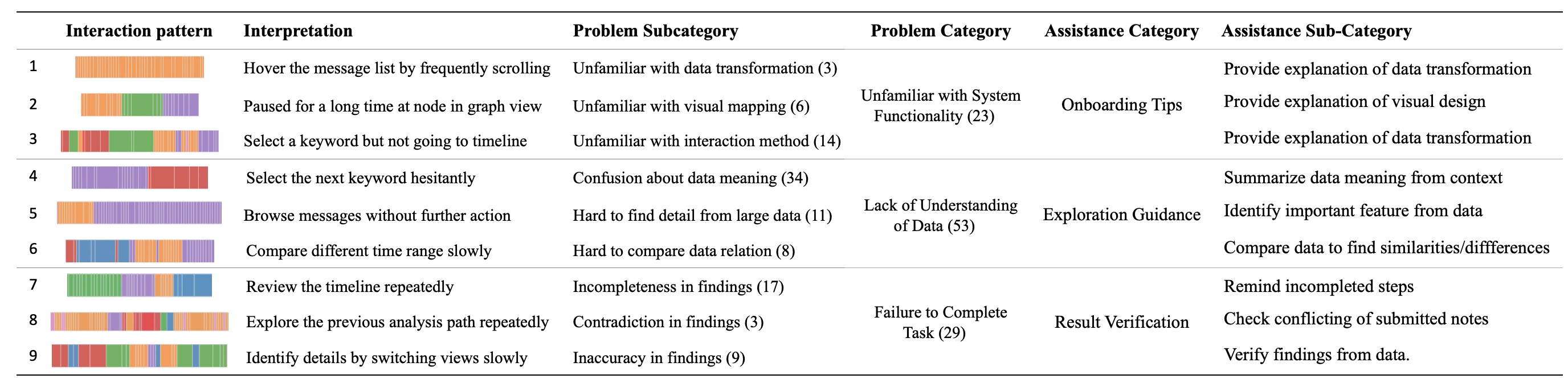}
    \caption{This table shows various user interaction patterns identified during the study, along with the interpretations, problem subcategories and categories. It also links these interaction behaviors to the appropriate assistance, based on the identified issues encountered by users.}
    
    \label{fig:patterns}
    \vspace{-10pt}
\end{figure*}

\section{Formative Study}
\label{sec:formative_study}

To answer three research questions (RQ1-RQ3) described in Sec.~\ref{sec:introduction}, we conducted a formative study observing users completing VA tasks.

\subsection{Study Set-up}

The study involved 10 postgraduate students (P1-P10), ranging in age from 24 to 29 years old. 7/10 participants have prior research experience in VA, covering education (P1, P9), disease transmission (P2), explainable AI (P10), social media topic analysis (P2, P4), and others have experience in using VA systems.
The system we used is designed for the 2021 IEEE VAST Challenge and awarded in Mini Challenge 3~\cite{Peng2021Mixed}. 
 The task is to \textit{identify as many public safety risk events with time, location, and characteristics as possible.}

\textbf{Procedure.} The overall session lasted around 1 hour, consisting of a think-aloud exploration and a follow-up interview.
The study began with a 5-minute briefing session where the participants’ demographic information was collected, including their age, major, data analysis experience, and familiarity with visualization tools. Following this, the background and functionality of the testing system were introduced.
Then, participants were invited to freely use the system to familiarize themselves with its features.
Next, we asked them to conduct a 15-minute think-aloud data exploration using the system.
After the exploration, participants were interviewed about when they felt stuck or uncertain and what proactive help they thought would have been useful.
The study is conducted through one-on-one meetings. 
We collected three kinds of data: user interaction logs, \revised{think-aloud data (video and transcripts)}, and follow-up interviews \revised{(notes)}.

\subsection{Data Collection}
% We describe the data collection process and the methods used to analyze the collected data. We collected user utterances and application-specific interaction data to analyze the semantic information of user interactions. 
% Building upon the taxonomy proposed by \cite{yi2007toward}, we identified three primary interaction categories:

% \begin{table}[htbp]
% \centering
% \caption{Interaction Types and Examples}
% \label{tab:interaction_taxonomy}
% \small
% \renewcommand{\arraystretch}{1.2}
% \begin{tabular}{p{1cm}p{3cm}p{3cm}}
% \hline
% \textbf{Type} & \textbf{Definition} & \textbf{Examples} \\
% \hline
% Select/ Filter & Choosing interesting elements to filter other views & Selecting a time range to update maps, graph, messages \\
% Abstract/ Elaborate & Showing an area of information more or less detailed & Hovering on a node to see the first-order neighbor node\\
% Encode & Changing visual representations to show different features & Switching y-axis from message type to event type  \\
% \hline
% \end{tabular}
% \end{table}

To support analysis of interaction behaviors, we adopt a structured parameter framework capturing key aspects of each interaction. Each event is described by attributes such as \textit{actionType} (e.g., click, brush, filter), \textit{view} (the visual context), \textit{element} (the visual mark), and the associated \textit{data}. We also log the precise \textit{clickTime}, and \textit{thinkTime} which reflects the pause between actions or duration of sustained interactions, indicating user engagement.

\begin{figure}[h]
    \centering
    \includegraphics[width=\linewidth]{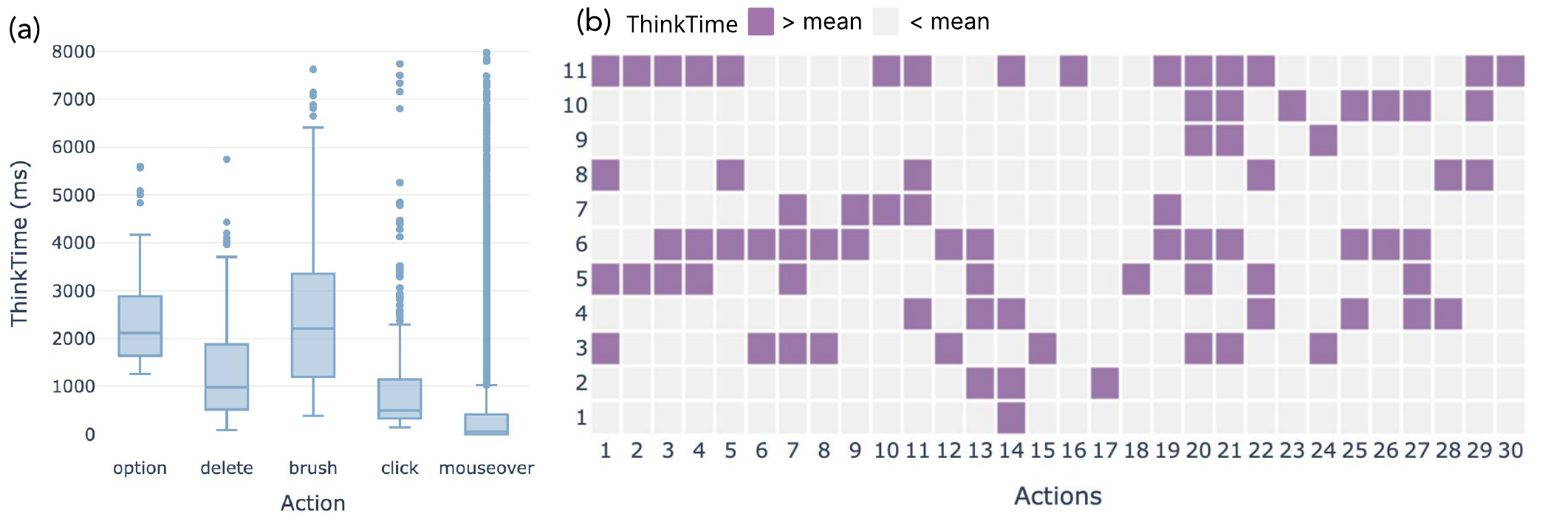}
    \caption{
        \revised{Visualization of user thinking time during analysis.}
        (a) Box plot shows distribution of \textit{ThinkTime} (in seconds) for different action types, with outliers indicating variability in decision-making. (b) A heatmap shows interactions across 30 steps, with each dark square indicating a longer thinking time than that type of action's average thinking time.
    }
    \label{fig:behavior-analysis}
    \vspace{-10pt}
\end{figure}

\begin{figure*}[t]
    \centering
    \includegraphics[width=0.91\linewidth]{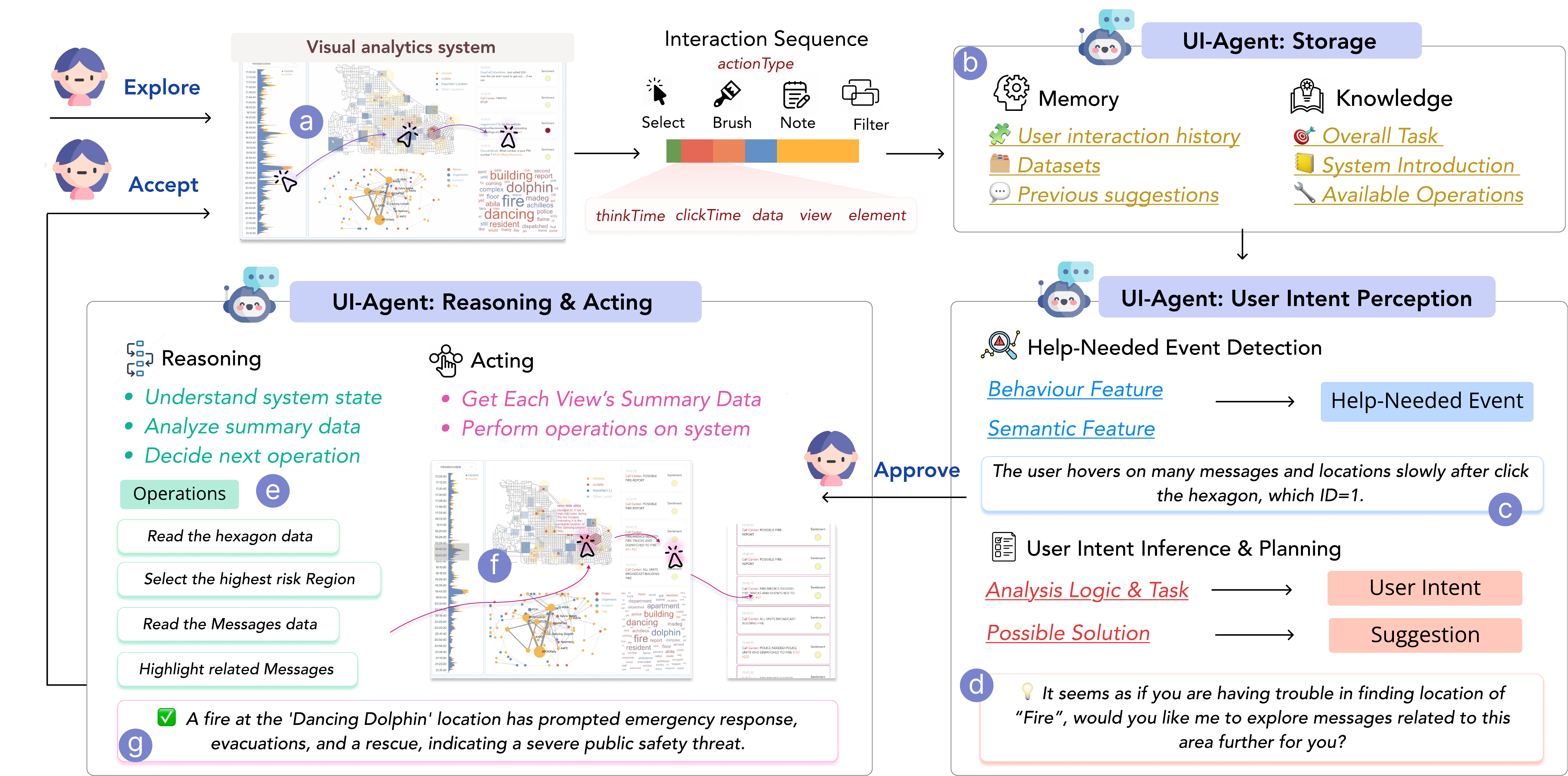}
    \caption{The workflow of ProactiveVA. \revised{The agent is responsible for detecting help-needed events (c), inferring user intent and providing suggestions (d), analyzing data to decide the next operation (e), performing actions on the interface (f), and storing history to facilitate the collaborative loop (b). The user explores the system (a), reviews and adjusts the agent's output (d), and controls the level of proactivity to ensure effective support (g).}}
    \label{fig:workflow}
    \vspace{-10pt}
\end{figure*}

We retained a total of 9,216 interaction records, averaging 916.7 per participant.
Prior work by Battle et al. \cite{BattleCharacterizing2019} observed that VA often follows a ``bursty'' interaction pattern, where users alternate between planning phases (longer pauses) and execution bursts (rapid actions).
Inspired by this, we examined the distribution of thinkTime in our data to identify potential moments of user hesitation or task transitions.
As shown in Fig.~\ref{fig:behavior-analysis}, the average thinkTime is around 2–3 seconds, and sequences of extended pauses frequently occur.
These extended pauses indicate potential user hesitation, and thus serve as a practical signal for activating the proactive agent.

% We analyzed the dataset both qualitatively and quantitatively. The goal of the analysis is to derive objective insights into user behaviors, task performance, and interaction patterns from the collected data. 

% \textbf{Task Performance.} 

\subsection{Data Analysis} 

To better understand the underlying reasons behind users' hesitations, we developed a visualization tool that aligns think-aloud data with interaction logs over time.
To investigate the reason behind the observed continuous pauses, we watch the videos and map think-aloud logs to the corresponding interaction sequences.
Then, two authors independently coded the logs, videos, and think-aloud data. We classify interaction sequences for all participants using the tool. 
The codes were refined through multiple rounds of discussions with co-authors, such as errors, omissions, confusions, etc, referencing prior research on proactive agents for data analysis \cite{TabalbaArticulatePro2024}, coding \cite{ChenNeed2024} and data wrangling \cite{HeerInteractive2012}. 
Finally, we obtained a total of 105 problems and corresponding assistance needed and intervention requirements from participants' feedback. The findings are detailed in the following sections.

\subsubsection{RQ1: When Do Users Need Proactive Help?}

In addition to consider \textit{thinkTime} as a fundamental indicator, we also summarized specific scenarios where users typically face problems. These problems relate primarily to the three stages of VA that the user is in, which is aligned with previous study~\cite{ZhaoLEVA2024}.

\begin{itemize}[leftmargin=*]
    \item \textbf{Unfamiliar with system functionality (23/105).} The majority of users experience confusion due to their (1) unfamiliarity with the system's interaction, which often leads to hesitation~(Fig.~\ref{fig:patterns}-3, 14/23). P4 shared, \textit{``At first, I wasn’t sure what I could click on and what I couldn’t. I kept trying things, but nothing happened.''}
    Moreover, the uncertainty is also reflected in (2) unfamiliar with understanding the system's visual encoding~(Fig.\ref{fig:patterns}-2, 6/23) and (3) data algorithms~(Fig.\ref{fig:patterns}-1, 3/23). As P5 mentioned, \textit{``I don’t fully understand how the system presents the data visually. Sometimes, I feel like I’m missing the meaning behind the colors and shapes.''} 
    % This lack of understanding often leads users to spend extra time figuring out how the system works, thus slowing down their exploration.
    
    \item \textbf{Lack of understanding of data (53/105).} Users often feel (1) confused about data meaning~(Fig.\ref{fig:patterns}-4, 34/53). P1 mentioned, \textit{``It's hard to make sense of the nodes meaning without understanding the messages.''} The (2) large amount of data makes it challenging for users to distill meaningful insights quickly, leading to delays in progression~(Fig.\ref{fig:patterns}-5, 11/53). P9 shared, \textit{``There's just so much data, I can't focus on what really matters.''} 
    (3) Comparing data across different views imposes a cognitive load.~(Fig.\ref{fig:patterns}-6, 8/53). P10 noted, \textit{``Switching between views and trying to link things together is exhausting. I often forget key details when I’m jumping between different views.''}
    
    \item \textbf{Failure to complete task (29/105).} Users often (1) forget details, requiring them to backtrack~(Fig.\ref{fig:patterns}-7, 17/29). P1 reflected, \textit{``I often forget the location or time of the messages, and I need to go back and recheck it, which slows me down.''} 
    Users often arrive at (2) incorrect conclusions but fail to recognize the errors, leading to repeated mistakes and cycles of confusion~(Fig.\ref{fig:patterns}-9, 9/29). P9 expressed, \textit{``I thought I had correctly identified all the key points earlier, but I missed something and repeated the wrong conclusion without realizing it.''}
    Users sometimes fail to notice (3) contradictions in their findings~(Fig.\ref{fig:patterns}-8, 3/29). P5 shared: \textit{``Initially, I reported the fire on the left and the shooting on the right, but later I found it was wrong.''}
\end{itemize}

\subsubsection{RQ2: What Assistance Should Be Provided?}

Based on user feedback, the proactive agent should offer the following types of assistance (Fig.~\ref{fig:patterns}):

\begin{itemize}[leftmargin=*] 
\item \textbf{Tips for onboarding.} The system should provide tips when users are stuck, helping them continue their operation. P4 stated, \textit{``When I get stuck or unsure in a view, it would be helpful to give me tips on how to proceed or correct my course of action.''} This assistance is intended to guide users and prevent frustration during analysis.
\item \textbf{Guidance for exploration.} The agent should suggest related data based on task relevance. For example, P5 mentioned, \textit{``When I’m filtering data, the agent should help me explore related information or recommend new areas to check.''} This ensures users are encouraged to explore beyond their current view and discover related insights.
\item \textbf{Reminders for errors, conflicts and omissions.} The agent should provide reminders when users may have missed crucial details, such as when selecting time or location filters. P3 noted, \textit{``Sometimes, I forget to include details like exact location or time, and the system could help remind me if something is missing.''} This kind of reminder ensures that users don’t overlook critical aspects of their analysis.
\end{itemize}

\subsubsection{RQ3: How Should the Assistance Be Presented?}

Based on user feedback, the following are the requirements for how the proactive agent should be presented:

\begin{itemize}[leftmargin=*] 
\item \textbf{Subtle but accessible suggestion.} The presence of the agent should not interrupt the user’s workflow, but should provide assistance in a subtle format, such as a floating window. Users should be able to open or close these help boxes as needed. For example, P5 mentioned, \textit{``I would prefer it to be unobtrusive, but readily accessible.''}
\item \textbf{Combination of text and visuals.} When users are unfamiliar with certain data concepts, the agent should combine visuals and text to explain. For instance, P4 stated, \textit{``I would like it to explain the event relationships with both images and text, helping me understand complex concepts.''} This multimodal approach is especially beneficial for improving users’ understanding of complex analysis tasks.
\item \textbf{Step-by-step guidance.} Users expect the agent to guide them step-by-step when dealing with complex analysis. P3 noted, \textit{``It would be helpful if the agent could guide me progressively. After each action, I would like to how it will be explored further.''} This indicates a preference for gradual assistance, allowing users to follow through the process at their own pace.
\end{itemize}

\subsection{Design Requirements Analysis}
\label{sec:requirements}

Combining the formative study findings and existing literature in the field of visual analytics, we summarize the design requirements for the proactive agent (DR1-DR3):

\textbf{DR1: Perceiving user needs by combining \revised{behavioral and semantic features} (RQ1).} 
    To provide timely assistance, \revised{the system should extract both behavioral and semantic features from users’ interactions and notes.
    We define behavioral features as including} temporal features (e.g., prolonged pauses) and action features (e.g., repetitive toggling)~\cite{BattleCharacterizing2019}. These may indicate that the user has difficulty with system onboarding or data exploration~\cite{ GotzBehaviordriven2009}. \revised{Meanwhile, semantic features are derived from user-submitted notes, e.g., missing findings, contradictory or inaccuracy conclusions. This feature signals that the user needs reminders to correct the submitted note. 
    } 
    
    \textbf{DR2: Tailoring assistance based on problem categories (RQ2).} After identifying that user needs help, the proactive agent should infer the specific type of difficulty the user is facing. For onboarding issues, \revised{the agent should infer user intent either based on behavior features and the system’s functional knowledge}, to determine which particular function the user is struggling with, and to offer specific tips. For data exploration issues, \revised{the agent should infer the user’s analysis intent based on behavior features, data relationships and system workflow. To help user finish their analysis task, the agent should propose a suggestion}~\cite{StoiberPerspectives2022}. For errors or omissions, \revised{the agent should analyze the data tables to identify the missing or erroneous parts in the notes}~\cite{HarrisonJupyterLab2024}, ensuring a more accurate analysis.
    
    \textbf{DR3: Ensuring transparent, \revised{controllable}, and non-intrusive assistance (RQ3).} 
    To support user trust and autonomy, the proactive agent should provide assistance that is both transparent and controllable~\cite{musleh2025trustme}. Suggestions should be accompanied by clear explanations, combining visual and textual cues~\cite{ShinDrillboards2024} or progressively guiding users when faced with complex analysis tasks~\cite{perez2025coupling, StolperProgressive2014}. Users should be able to preview and evaluate suggestions before applying them~\cite{ChenNeed2024}, \revised{and adjust settings such as the frequency of assistance, types of support (e.g., guidance, verification), and whether the agent can initiate actions}~\cite{DengHumancentered2024a}. To minimize disruption, the agent should deliver help through non-intrusive interfaces such as tooltips or subtle visual markers~\cite{yanez2025state}.

\section{ProactiveVA}

The UI agent's unique advantage lies in its \textit{bidirectional capability to interpret interface-level interactions and programmatically manipulate interface}. 
Therefore, we propose a three-stage framework utilizing an LLM-based UI agent to perceive users' proactive needs and provide effective solutions through autonomous action execution.
\revised{The UI agent has some main components: a \textbf{perception} module for environment sensing, a \textbf{reasoning} module for decision-making, and an \textbf{acting} module for interacting with the system. 
The LLM serves as the ``brain'' of the agent for language understanding and generation, processing the information the agent receives.}
The overall process relies on two kinds of information in the storage: \textit{memory} and \textit{knowledge}~(Fig.~\ref{fig:workflow}b). The \textit{memory} stores the recent interaction history, dynamically updated dataset, and previous suggestions. 
\textit{Knowledge} includes the overall task, system introduction, and available operations. 
Our implementation and studies use the model DeepSeek-V3~\cite{liu2024deepseek}. We provide input and output examples of each stage in Appendix A.

\subsection{UI Agent: User Intent Perception}
\label{sec:perception}

\revised{The UI agent in our system needs to perceive human interaction for collaborative analysis. It not only observes user actions, but also interprets how those actions affect the system state, how visualizations relate to one another, and how data is encoded. Our two-step approach is designed with these VA characteristics in mind. The agent first identifies help-needed moments from interaction sequence and then speculates on the possible analytical intent of the user, plans the following steps.}

\textbf{Subtask 1: Help-needed event detection.} 
\revised{The detection of help-needed events in onboarding and exploration mainly relies on the behavioral features extracted from user interaction logs, while verification is more about checking the semantic features from notes (DR1). 

Firstly, the behavioral features are extracted from interactions. A common type is \textit{prolonged pause},} where a user pauses beyond a predefined ThinkTime threshold. 
\revised{Moreover, \textit{repetition} behavior offers a different clue, such as toggling filters back and forth, clicking the same region multiple times, or switching between views without making progress. 
This behavior implies that the user is actively exploring but lacks a clear strategy.
For example, during social media event analysis, the agent identifies that the sequence is slow and has repeated hovers~(Fig.~\ref{fig:patterns}c), which may suggest difficulty in understanding massive messages. 
The description of this sequence is summarized as a help-needed \schemaname{event}.
Secondly, the semantic features are extracted from user-authored notes, such as potentially \textit{inaccuracy or incompleteness} (e.g., claiming a region’s profitability increased when it actually declined). 
The agent should compare the note content with the system's data. If data facts are inconsistent, the user should be reminded.
To handle these three tasks, we employ parallel agents, where each agent independently monitors its own specific triggers. This allows the system to handle different types of assistance simultaneously.
We provide examples to the model as few-shot prompts, allowing it to infer over new features by analogical pattern matching (Fig.~\ref{fig:patterns})
}

\textbf{Subtask 2: Intent inference and planning.} Once an \textit{event} has been detected, the system must interpret what the user is trying to achieve and what kind of assistance would be most appropriate (DR2). 

\revised{During the \textit{onboarding} phase, based on the detected \schemaname{event}, the agent should further infer the reason and derive user intent. The agent should combine the system workflow and the functions complexity to infer whether the user is unfamiliar with visual encoding or the interaction and provide tips.
For example, if a user selects a keyword but does not proceed to the timeline view, the system may infer uncertainty about interaction logic and offer a \schemaname{suggestion} such as ``Click to see how this keyword evolves over time''.}
In the \textit{exploration} phase, the system analyzes the \schemaname{event} in relation to the visualization structure and data semantics, in order to infer more complex analytical \schemaname{intent} such as comparison, filtering, or trend detection.
\revised{For example, in Fig.~\ref{fig:workflow}d, the agent examines the content of the hovered messages and finds repeated mentions of ``Fire'' and infers that the user may be trying to find the detailed location of the Fire event and finally generate a \schemaname{suggestion} to ask if the user need its help to complete the analysis.
In the \textit{verification} phase, as the help-needed event already indicates specific errors or omissions in the notes, the agent does not need to infer user intent but generates rule-based suggestions for revision. Each \schemaname{suggestion} includes the type of issue (e.g., factual error, internal conflict, or task omission), a brief comment explaining the problem, a corrected answer, and keywords from the original text to highlight the affected parts.
} 

\begin{figure*}[t]
    \centering
    \includegraphics[width=\linewidth]{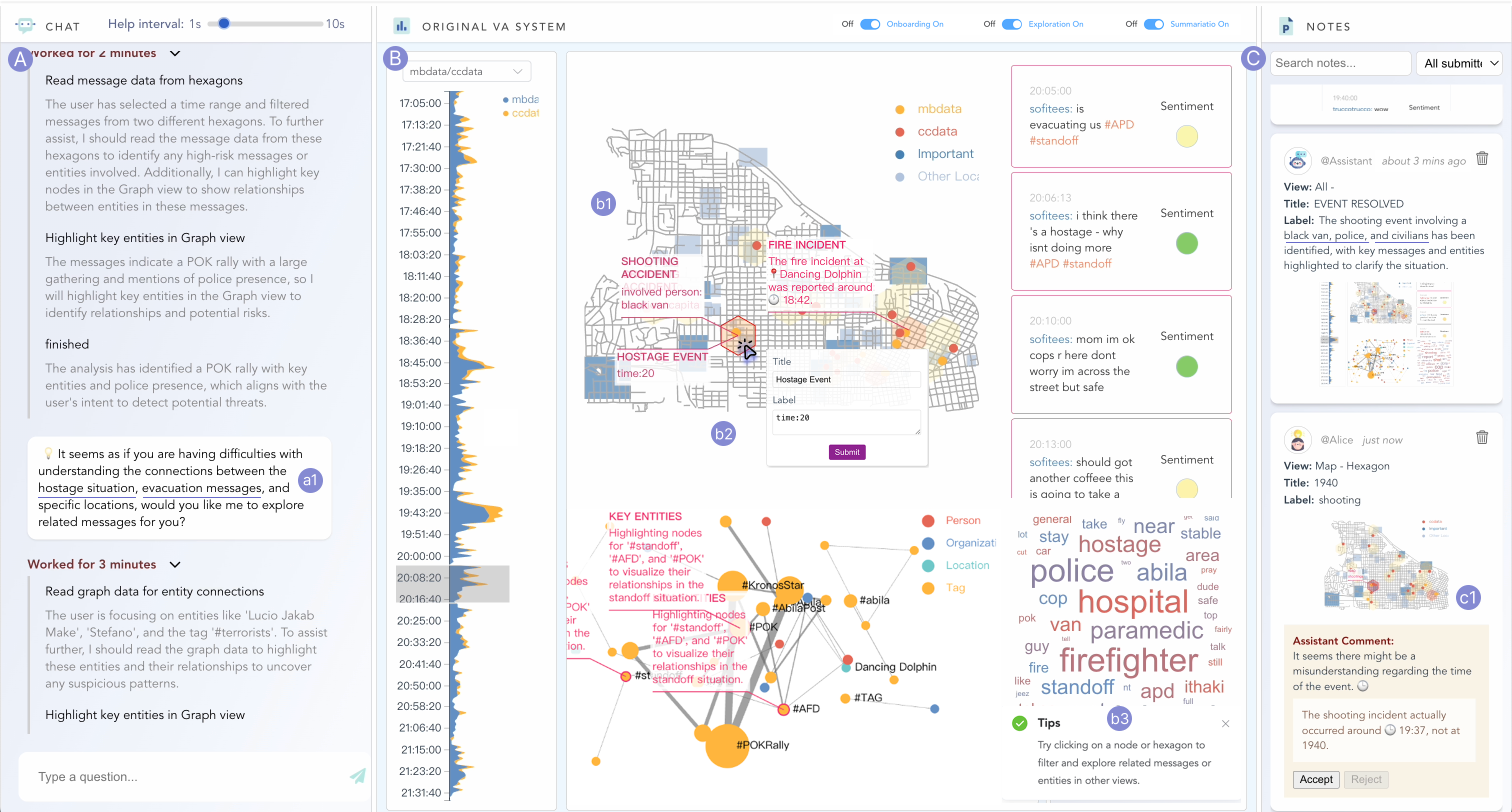}
    \caption{
        The overview interface of our UI agent-assisted system:
        (A) Chat View displays proactive suggestions from the agent alongside its reasoning process (a1), helping users understand the system’s analysis.
        (B) VA interface serves as the primary workspace for human-agent interactive exploration.
        Key findings identified by the agent are highlighted within the visualization (b1), enabling users to visually link insights with data points. Users can also add custom annotations (b2) and access onboarding tips (b3) for guidance.
        (C) Notes View stores both the user’s and agent’s findings for review and summarization.
        The agent can also validate the user’s notes to improve accuracy (c1).
    }
    \label{fig:interface}
\end{figure*}

\subsection{UI Agent: Interactive Reasoning \& Acting}
\label{sec:reasoing_acting}

This stage involves reasoning and acting to interactively explore the system and provide insights for users. Inspired by the ReAct~\cite{YaoReAct2023}, our process involves breaking down the initial plan into smaller sub-tasks and determining the next step in an iterative manner. After each action, the agent reflects on the current state and generates a \textit{thought} that helps inform the next operation. This framework allows the agent to autonomously debug and adjust actions for the VA system, providing transparent and explainable assistance throughout the process (DR3).

\textbf{Subtask 1: Reasoning.}  
To execute the initial \schemaname{plan}, the UI agent must interpret and reason over data distributed across different types of components, including visualizations, texts (e.g., key indicators), and UI widgets (e.g., filters)~\cite{SrinivasanDashboard2024}. Each component provides a distinct lens on the data and must be understood in context.
\revised{Based on the system’s visual structure, the agent can use a set of tools, \textit{readData}, \textit{select}, and \textit{filter}, to get data, apply selections, or adjust filters in views.
For example, the agent may start by reading data from the hexagon grid to assess regional risk levels (Fig.~\ref{fig:workflow}e). It identifies a high-risk region and selects it. Then, it retrieves and analyzes associated message content, where repeated mentions of ``fire'', ``rescue'', and the location “Dancing Dolphin” are found. Based on this evidence, the agent generates a \schemaname{finding} such as: \textit{``A fire at the `Dancing Dolphin' location has prompted emergency response, evacuations, and a rescue, indicating a severe public safety threat''} (Fig.~\ref{fig:workflow}g).
Each reasoning step includes a \schemaname{thought} to interpret the current state and a candidate \schemaname{operation} for the next action. This loop continues until the analytical goal is satisfied or the agent determines that no further reasoning is needed.}

\textbf{Subtask 2: Acting.}
After generating \schemaname{operation}, the action module executes this operation to change the environment state. The action module not only performs the operation but also monitors the system’s response, providing \schemaname{feedback} to the reasoning module (Fig.~\ref{fig:workflow}f). This feedback includes error reports, system state changes, or any inconsistencies encountered in the environment. For instance, if the system encounters an error, such as incorrect data being processed or a failure in visual mapping, the action module will notify the reasoning module to adjust its strategy accordingly.

\subsection{User Interface}

We designed an interface to bridge the LLM-based UI agent, the VA interface, and users. \revised{To support transparent, controllable, and non-intrusive assistance (DR3),} the interface consists of three components, which work together to assist users throughout the visual analysis process, as shown in Fig.~\ref{fig:interface}.

\textbf{Chat view.}
When the agent detects that the user needs help, it offers assistance and suggestions for further exploration in the Chat View. If the user agrees, the agent takes over the workflow and continues the exploration until the task is resolved. During this process, the agent presents its next steps and thought process in real-time in the Chat View (Fig.\ref{fig:interface}A). Additionally, findings explored by the agent are presented within visualization annotations and highlights in the Original System View (Fig.\ref{fig:interface}B), each including a title and label to clearly indicate where the agent has identified findings. These findings are also stored as note cards in the Notes View (Fig.\ref{fig:interface}C).

\textbf{Visual analytics interface.}
The VA interface is the central component, allowing users to independently explore the system (Fig.\ref{fig:interface}B). While interacting with the system, the agent monitors user behavior and proposes suggestions. System-level tips appear in the bottom-right corner of the window and disappear after 5 seconds if not interacted with, ensuring minimal disruption. The interface also includes a frequency control slider for proactive interventions and switches for onboarding tips, exploration guidance, and verification reminders, allowing users to customize their experience.

\textbf{Notes view.}
The Notes View (Fig.\ref{fig:interface}C) supports the summarization process during analysis. Both the agent and the user can submit notes. If the agent detects errors, contradictions, or omissions in the user’s notes, it will provide feedback. Users can apply the agent's suggestions, ensuring collaboration remains non-intrusive while promoting cooperative analysis.

% 我们设计了一个交互界面来支持section~\ref{sec:workflow}中的pipeline，方便用户理解和评估建议（DR4)。

\section{Evaluation}
The evaluation of ProactiveVA includes algorithm performance assessment, a case and expert study, and a user study to comprehensively evaluate the system.

\subsection{Algorithm Evaluation}
\label{sec:experiment}

\begin{table*}[ht]
    \centering
    \caption{
        Evaluation of model performance across different data analysis tasks and metrics.
    }
    \setlength{\aboverulesep}{0pt}
    \setlength{\belowrulesep}{0pt}
    \vspace{-7pt}
    \renewcommand{\arraystretch}{1.12}
    \begin{tabular}{l|c|c|c|c|c|c|c|c|c}
        \toprule
        \multirow{2}{*}{Category} & \multirow{2}{*}{Count} & Time (s) & Step Count & \multicolumn{3}{c|}{LLM Evaluation} & \multicolumn{3}{c}{\revised{Human Evaluation}} \\
        \cline{5-10}
        & & Mean (±Std) & Mean (±Std) & Task Com. & Data Acc. & Path Eff. & Task Com. & Data Acc. & Path Eff. \\
        \midrule
        Total/Avg.      & 100 & 35.94 (±13.47)  & 4.36 (±1.61)  & 4.15  & 4.96  & 4.45  & 4.38 & 4.85 & 4.67 \\
        Comparison  & 17  & 40.81 (±10.44)  & 5.12 (±1.41)  & 4.06  & 5.00  & 4.24  & 4.41 & 4.94 & 4.65 \\
        Trend       & 20  & 31.90 (±9.49)   & 3.80 (±1.03)  & 4.20  & 5.00  & 4.70  & 4.45 & 4.85 & 4.80 \\
        Performance & 31  & 33.72 (±8.90)   & 3.94 (±0.88)  & 4.16  & 4.87  & 4.52  & 4.42 & 4.81 & 4.71 \\
        Correlation & 11  & 33.81 (±11.77)  & 4.36 (±1.77)  & 3.64  & 5.00  & 4.18  & 3.73 & 4.64 & 4.27 \\
        Dimension   & 21  & 40.22 (±20.93)  & 4.90 (±2.37)  & 4.43  & 5.00  & 4.43  & 4.57 & 4.95 & 4.71 \\
        \bottomrule
    \end{tabular}
    \label{tab:task_metrics}
    \vspace{-10pt}
\end{table*}

\revised{This section presents a batch simulation evaluation of UI agent's reasoning and acting capabilities for interactive exploration by itself (Section~\ref{sec:reasoing_acting})}.
The evaluation was performed on a profitability analysis system used for Case study 2 (Section~\ref{sec:expert_inter}).
The experiment comprises three phases: task generation, automated sandbox execution, and subjective scoring, with an analysis of system performance.

\revised{\textbf{Method.}} Given the challenges in collecting real user interaction data and tasks as ground truth, we employed DeepSeek R1 to generate 100 tasks and evaluate the agent's performance. 
The input tasks are five categories:
(1) Comparison analysis, 
(2) Trend analysis, 
(3) Performance analysis, 
(4) Correlation analysis \revised{(e.g, explore relationships between variables such as discounts, sales, profit}, 
(5) Dimension-specific exploration \revised{(e.g, drill into specific data columns or multi-dimensional combinations)}.
\revised{Given a task, the agent will output thoughts, actions, and insights.}
We develop a sandbox testing system that sequentially injects each task into the system. The system then autonomously executes analysis, tool invocation, task execution, and data fact presentation. We record inputs, outputs, execution time, and step count for further evaluation.
\revised{We employ DeepSeek R1 to assess the system’s performance and have two data analysis experts evaluate with the same criteria.}
The evaluation is based on three key factors (scored on a scale from 0 to 5): \revised{(1) Task completion (correct data coverage and task type)
(2) Data accuracy (no miscalculation)
(3) Path efficiency (no repetitive actions).}

\revised{\textbf{Results.}} As shown in Table \ref{tab:task_metrics}, the system demonstrated robust performance with an efficient execution and high-scoring evaluation results. 
From the perspective of different task types, we can observe that Comparison Analysis may require more operations, while Correlation \& Impact Analysis may be harder to get a good result. 
\revised{When comparing the LLM's evaluation to human evaluation, the LLM's average score is 4.52, while the human average is 4.63. This close score indicates that the LLM, as an automated evaluation mechanism, is a feasible option. However, differences in evaluation emerge in specific tasks and dimensions. Notably, the LLM assigns lower scores for Task Completion (4.15 vs. human's 4.38), primarily because it struggles to interpret visual outputs (e.g., data highlighting) and may criticize insights as ``not profound.'' In contrast, humans rely on these visual cues. On the other hand, the LLM rates Data Accuracy higher (4.96 vs. humans' 4.85), which can be attributed to the LLM raters' ability to capture fine-grained data details, leading to greater confidence in its evaluations.}
Overall, the results demonstrate our system's effectiveness in handling diverse analytical tasks. More assessment details are provided in Appendix B.

\subsection{Case Study and Expert Study}
\label{sec:cases}

In order to examine the effectiveness of our system, we implemented ProactiveVA in two applications.
One is the social event analysis system~\cite{Peng2021Mixed} introduced in Section~\ref{sec:formative_study}.
The other is a classic profitability analysis dashboard from Tableau~\cite{Superstore_embedded}.
% 为了验证我们系统的有效性，我们把proactivVA接入到两个applications中，一个是section3中用到的social event analysis system，另一个是来自tableau的一个经典的sales analysis dashboard。

\begin{figure*}[ht]
    \centering
    \includegraphics[width=\linewidth]{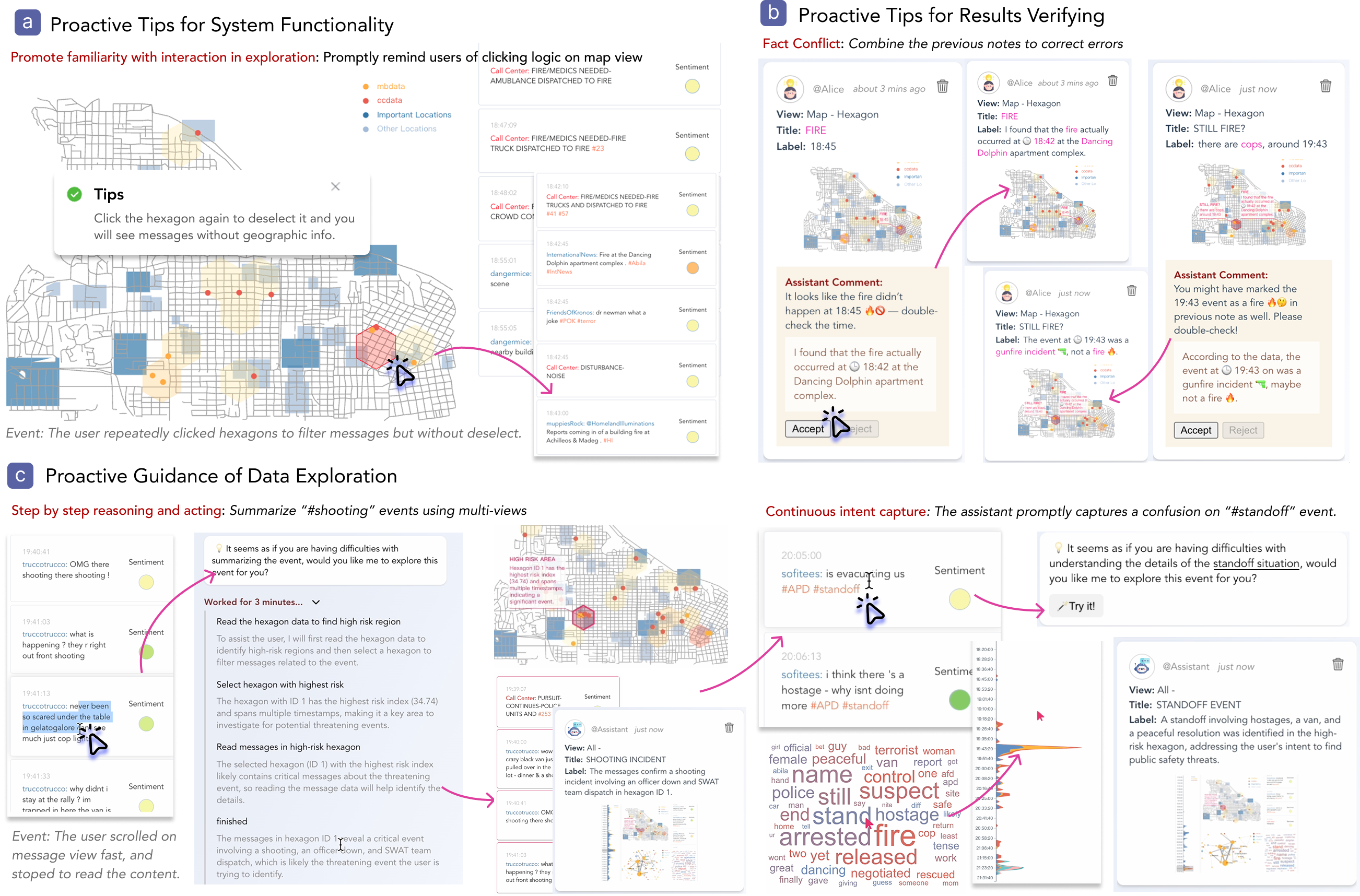}
    \caption{Three cases of proactive agent interventions: (a) onboarding guidance for data filtering, (b) error correction in event analysis, and (c) interactive exploration of a shooting incident. Each demonstrates context-aware assistance in addressing different analytical challenges.}
    \label{fig:case}
    \vspace{-10pt}
\end{figure*}

\subsubsection{Case 1: High Risk Event Analysis}
\revised{This case study aims to evaluate the effectiveness of our method in handling multimodal data (e.g., spatial-temporal and texts). The study was conducted by an expert from our author team with 6 years of data analysis experience, including over 2 years in developing and using VA systems, particularly in graph data visualization and abnormal event detection. The analysis goal is to identify high-risk public safety events.}

\textbf{Proactive tips for system onboarding.}
At the beginning, the expert first filtered a prominent peak (18:40–18:50) and then selected a geographic region on the map to examine related messages in the Message View (Fig.~\ref{fig:case}a). 
However, the initial filtering yielded few messages, prompting her to switch regions in search of actionable insights repeatedly.
% \yz{Why is this a sign of confusion?}
The agent then intervened with a contextual tip:
\textit{``Click the hexagon again to deselect it. This will show messages without geographic tags.''}
This guidance led the expert to realize that many messages (e.g., microblogs) lacked geographic metadata.
Upon deselecting the hexagon, the expert successfully uncovered critical reports of a building fire previously hidden due to over-filtering.

\textbf{Proactive guidance of data exploration.}
While analyzing a potential shooting event, the expert struggled to synthesize fragmented messages (e.g., \textit{``OMG there shooting!'', ``tramped in here the van is''}) amid rapid scrolling(Fig.~\ref{fig:case}c). The UI agent detected dilemma and intervened: \textit{``It seems you’re having trouble summarizing this event. Would you like me to help?''} Upon confirmation, it transparently executed a step-by-step analysis: (1) spatial filtering identified Hexagon ID 1 as the highest-risk zone (risk index=34.74); (2) temporal analysis isolated critical messages between 19:40-19:45; and (3) semantic parsing yielded the summary: \textit{``Shooting involving officer down and SWAT dispatch.''}

\textbf{Proactive tips for results verifying.}
In another scenario, the expert annotated a timeline note to mark the fire event’s start time as 18:45.
The proactive agent, leveraging real-time data validation, detected a discrepancy: the earliest fire-related messages actually appeared at 18:42. Upon reviewing the evidence, Bob accepted the suggestion and adjusted his analysis (Fig.~\ref{fig:case}b).
Later, when the expert identified a high-risk region around 19:45 based on keywords like ``fire'' and ``cops'', the agent flagged potential misinterpretation: \textit{``According to the data, the event at 19:43 was a gunfire incident, maybe not a fire.''} This helped the expert revisit the data and avoid a mistaken conclusion.

\revised{To sum up, this case demonstrates that the proactive UI agent can assist in onboarding by resolving misunderstandings, supports result verification through fact-checking, and guides exploration by bridging fragmented information. 
These interventions not only improved task efficiency but also helped prevent common analysis pitfalls such as over-filtering and semantic misinterpretation.
Moreover, the expert suggests that providing multiple diversified recommendation options may support more flexible and user-driven exploration.}

\subsubsection{Case 2: Sales Data Analysis}
\label{sec:expert_inter}

We implement our framework on a Tableau Public dashboard to evaluate its effectiveness in real-world business analysis scenarios. This implementation follows the methodology in Section~4, with adaptations made to accommodate Tableau's ecosystem. Besides interpretation and selection of charts, we also incorporate filter panels and key performance indicators, which are common in analytical dashboards~\cite{SrinivasanDashboard2024}. We leverage Tableau's Embedding API~\cite{Tableau2003Embedding} to enable bidirectional interaction between the agent and dashboard components. In this case, we use a classic profit analysis dashboard~\cite{Superstore_embedded}. 

\revised{We invited a domain expert who specializes in business analysis from a global top fast-moving consumer goods (FMCG) company. The expert holds a master's degree in data science and has 5 years of experience in dashboard-based analysis and uses it almost daily.}
The interview lasted around 1 hour with three phases: background survey and system introduction (15 minutes), free exploration (20 minutes), and interview (20 minutes).
\revised{The goal is to examine profitability differences and identify factors contributing to underperforming areas.}

\begin{figure}
    \centering
    \includegraphics[width=\linewidth]{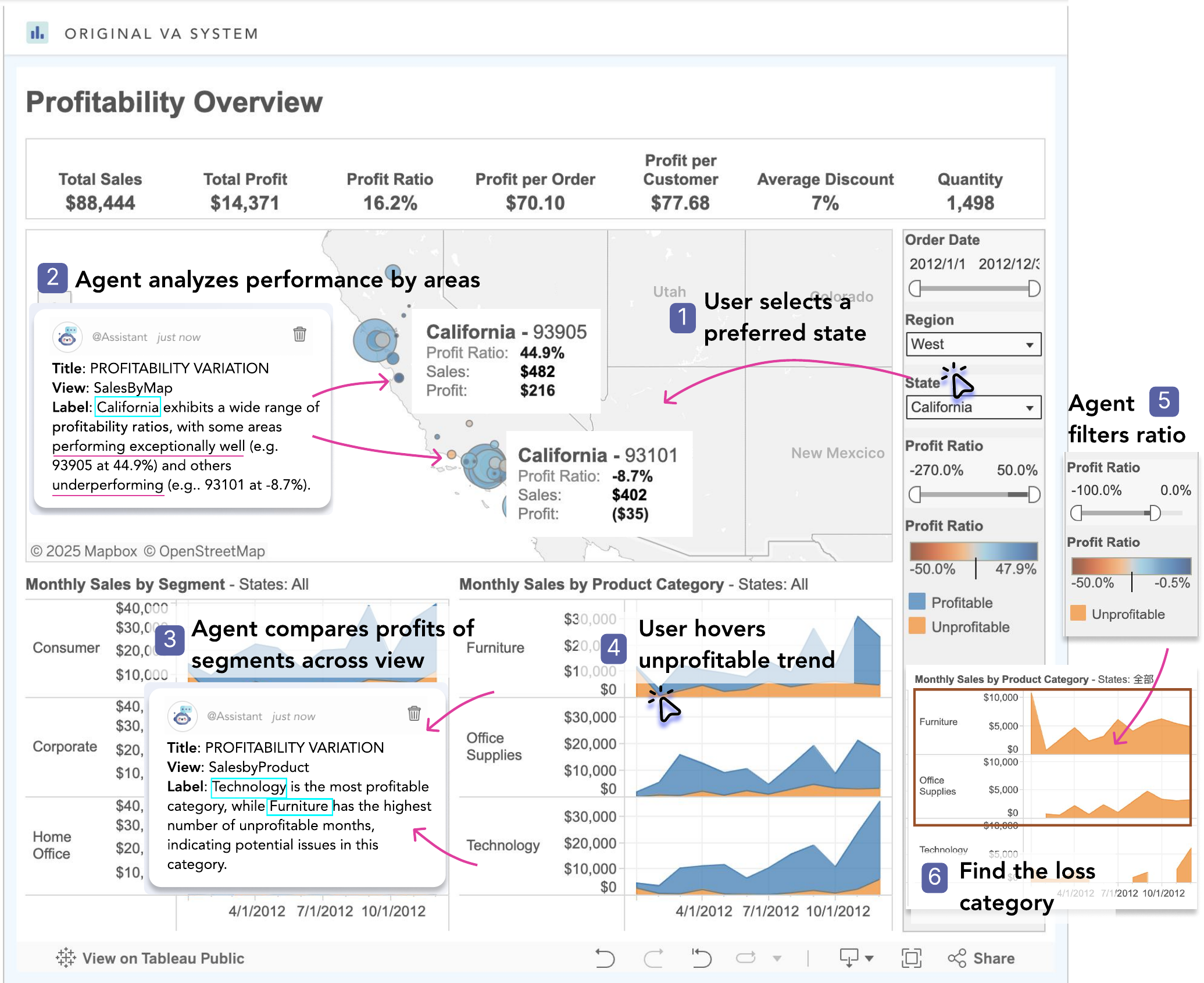}
    \caption{Case 2: Integrating ProactiveVA into Tableau Sales Data Analysis Dashboard. The expert and agent collaborate to identify unprofitable products. The expert decides on the direction of exploration, while the agent proactively offers suggestions and utilizes the dashboard features for interactive analysis.}
    \label{fig:case_tableau}
    \vspace{-10pt}
\end{figure}

\textbf{Promptly perception and insight surfacing}. The expert began an exploratory analysis by hovering over different map points and selecting California in the filter panel~(Fig.\ref{fig:case_tableau}-1). Then the agent proactively suggested analyzing geographic profitability trends: \textit{``It appears you're examining profitability ratios. Would you like to focus on specific ranges (e.g., highly profitable vs. unprofitable regions)?''} The expert agreed that \textit{``This was a timely suggestion, as profit ratio is an important factor in analyzing the reasons behind sales differences.''} The agent identified significant profitability variations in California, identifying both top-performing (93905 at 44.9\%) and underperforming (93101 at -8.7\%) areas~(Fig.\ref{fig:case_tableau}-2). The expert verified these insights on the map, stating, \textit{``It reminds me of an important finding. I have not noticed that these two small points had the greatest profit ratio difference.''} In addition, the agent also analyzes the trend and product category, proving its ability to understand different data~(Fig.\ref{fig:case_tableau}-3).

\textbf{Domain-specific operational suggestions}. During the analysis, the agent demonstrated its ability to operate the dashboard and presented professional analytical insights. For example, when the expert was focused on the unprofitable trend~(Fig.\ref{fig:case_tableau}-4), the agent proactively suggested, \textit{``Filter negative profit ratios to focus on underperforming areas.''} Using the filter panel, the agent adjusted the profit ratio slider to range from -100\% to 0, updating the entire interface~(Fig.\ref{fig:case_tableau}-5). Based on this, the agent identified differences in product categories and explained that California's low profit was primarily due to Furniture and Office Supplies~(Fig.\ref{fig:case_tableau}-6). The expert commented that \textit{``Yes, the filter for profit ratio is correct. These negative values should be prioritized. While our annual goals are typically growth-oriented, the actual analysis often focuses on a specific range below the target.''}

\revised{Overall,} during the analysis, the agent demonstrated notable analysis capability, identifying 12 actionable insights evaluated by an expert. 
In the process of analysis, no interactive error was reported, and the number of reasoning steps triggered ranged from 2-5, covering various insight types, e.g., comparison, trend, extreme, category analysis. 
\revised{The expert noted that the agent’s suggestions aligned well with real-world business logic and helped surface insights that might otherwise be overlooked. 
She also found that agent-led exploration was particularly helpful during cold-start analysis, and suggested that future iterations could incorporate new data and charts for dynamic task-driven analysis.
}

% 我们把主动式ui助手接入到了tableau public dashboard中，因为tableau有大量在线资源，是许多人日常进行业务分析的工具。具体来说，我们结合了tableau embedding api来让agent能够监听用户的交互，包括视图级别的hover和filter，同时让agent发起select交互，我们给了允许llms理解key indicator view，chart view and filters panel view，并在chart和panel上进行交互选择。详细的prompt和available interaction tools for agent 会放在appendix中。

\subsection{User study}
\label{sec:user_study}

We conducted an in-lab user study with participants of varying VA expertise to evaluate the effectiveness of ProactiveVA.

\textbf{Participants}. We recruited 10 users, \revised{aged between 23 and 27, with 5 males and 5 females, labeled as T1 to T10. They were postgraduate students majoring in data science (including undergraduate studies). Their data analysis experience ranged from 3 to 6 years. Among them, T1, T2, and T4 had 3-6 months of experience in VA, while the others had 2 years. Their specialized domains included financial data (T7), spatiotemporal data (T5), network and text data (T3), and event analysis (T9). In addition, except for T2 and T3, all other participants had research experience involving the combination of large models and human-computer interaction. All participants received \$8 as compensation.} \revised{The participants were selected to cover a range of experience levels and data domains, ensuring both analytical proficiency and relevance to the target user group of proactive visual analysis systems.}

\textbf{Task}. We asked participants to perform the same open-ended event analysis task as described in Section~\ref{sec:formative_study}. Participants were asked to explore the data for 15 minutes and identify as many events, times, places, and people as possible. This was aimed to compare with the formative study and assess whether the inclusion of a proactive agent would enhance or interfere with the users' analysis results.

\textbf{Procedure}. All studies were conducted in one-on-one offline sessions. Before the user study, we briefed the participants on the procedure and obtained their consent to record the study. During the study, we first introduced the various components of our system and explained the relevant interactions. Participants were then guided through the onboarding process to understand the features of the original VA system. Once they confirmed their understanding, they entered the exploration phase, which lasted for 15 minutes. At the end of the session, we interviewed the participants about their experiences. They were also asked to complete a 7-point Likert questionnaire to assess the effectiveness and usability of the system. In the questionnaire, 1 point indicated ``strongly disagree'', and 7 points indicated ``strongly agree''. The eleven effectiveness-related questions are shown in Fig.~\ref{fig:user-study}.
Questions Q1 to Q3 assess timeliness and proactivity, Q4 to Q5 focus on analytical capabilities and interpretability, Q6 to Q8 evaluate the system's practicality, and Q9 investigates the overall influence of proactive assistance during analytical tasks. It took around 20 minutes to finish the interview and the questionnaire. The whole user study lasted about 1 hour. Each participant received \$8 as compensation. The authors took notes to record feedback during the study.

% gyc、tianrui、wanyi、yihang、yifei、qiutian、xiaowen、liwen、ziyue、qianhui

\subsubsection{Feedback \& Results}
\begin{figure}
    \centering
    \includegraphics[width=\linewidth]{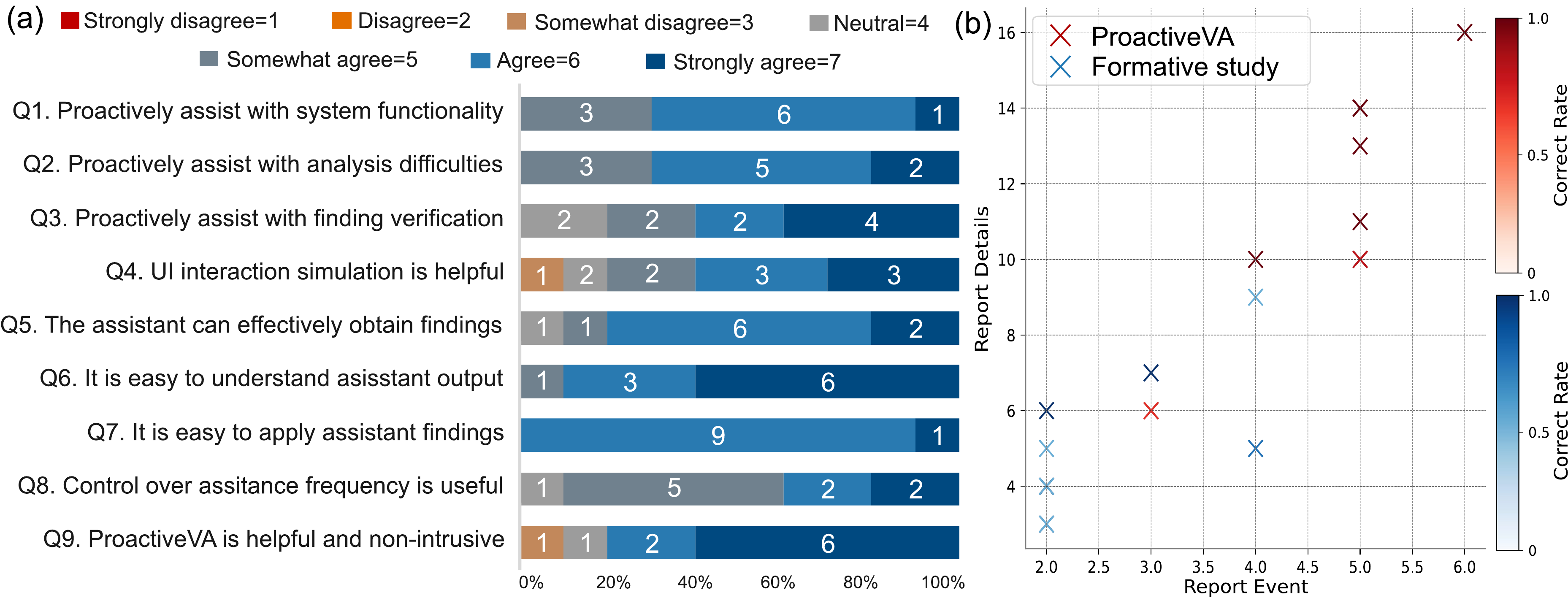}
    \caption{The results of quantitative evaluation of the effectiveness. (a) Subjective ratings of our system across various dimensions (Q1 to Q9). (b) Comparison between ProactiveVA and formative study, showing better performance in number of report events, report details and correct rate.}
    \label{fig:user-study}
    \vspace{-15pt}
\end{figure}

Overall, ProactiveVA received an enthusiastic reaction from participants. As shown in Fig.~\ref{fig:user-study} (a), participants deemed the assistance from the proactive agent and the system to be useful. In the following text, we summarize participants’ feedback on ProactiveVA.

% In this section, we report the quantitative and qualitative results of the user study. The quantitative results of our user study reflect participants’ ratings on both effectiveness and usability. The effectiveness ratings are shown in Fig.~\ref{fig:user-study} (a).

\textbf{Effective timeliness and proactivity (DR1).} As the results indicate, most of the participants felt satisfied with ProactiveVA. The proactive assistant demonstrated effective demand awareness, with users generally agreeing that it could promptly perceive their needs ($\mu=5.83, \sigma=0.87$). Participants particularly praised its real-time intent recognition for system functionality (Q1) and analytical exploration (Q2), as exemplified by T9's comment: \textit{``It conveniently understands my analysis intent based on where I have been.''} The agent' recommended suggestion also received positive feedback, with T3 noting: \textit{``The assistant not only identifies my current needs but also guides and automates subsequent tasks.''}

\textbf{Enhanced analytical performance (DR2).} We compared the task completion performance with the formative study (Fig.~\ref{fig:user-study} (b)). The result shows that with proactive assistance, users' analysis improved in terms of breadth, depth, and accuracy. In terms of analysis breadth, without assistance, the average number of reported events per person was about 2, whereas with assistance, it increased to about 4 per person. In terms of depth, the average number of reported details per person increased from 5 to 11. T6 remarked that \textit{``The agent can quickly summarize main points to supplement my understanding, saving my time to read one by one.}
In terms of accuracy, it improved from 0.78 to 0.95. 
T3 indicated that: \textit{``The notes comments are useful, which promptly correct my misinterpretations of the data.''}

\textbf{Improved understanding of agent behavior (DR3).}
Users reported that the system's preview function, step-by-step annotations, and self-explanation features significantly improved their comprehension of agent behavior and increased trust in exploratory results (Q6-Q7 $\mu=6.3, \sigma=0.64$). T10 commented, \textit{``I like these annotations and screenshots with notes, they let me clearly know how agent thinks and operates the system step by step.''}

% \textbf{增强人对agent的理解} 用户认为系统的preview，step by step 的annotations 和self-explanation可以极大的增强用户对agent行为的理解，增加对探索结果的信任，q6
% -q7的平均得分6.3，方差0.64。p8表示 p6 p7 agent自动笔记便于理解，p7和p8及时记录的笔记能沉淀思考
% p10表示“ui agent在系统上走过，留下高亮这个我很喜欢，可以让我清晰地看到它的思路。” 

\textbf{Facilitated human-AI collaborative analysis.} As shown in Fig.~\ref{fig:user-study} (a), most users did not deem ProactiveVA brought intrusive feelings ($\mu=6.1, \sigma=1.45$) and indicated that proactive analysis stimulated deeper thinking.
First, we observed that during initial exploration, when users were less familiar with the data and system, the agent played a more active role in guiding users and inspiring exploratory ideas. When analyzing user-agent interactions, we find that each user incorporated an average of 8 LLM-generated suggestions into their notes ($\sigma=3.59$). T10 added, \textit{``After reading its summaries, I knew which keywords to explore next. It advanced my thinking.''}
Second, as users gained proficiency, users became more active. We found that beyond the agent's notes, users also recorded an average of 5 additional notes. T6 exemplified this evolution: \textit{``Initially I followed its guidance due to system unfamiliarity, but later our roles reversed, I have more ideas to lead the exploration while the agent assists me to summarize data.''} Notably, T6's workflow showed effective collaboration, adopting 15 agent suggestions while submitting 10 notes, ultimately reporting 5 events and 11 detailed findings. Overall, all participants appreciated the ``user-centered proactive assistance'' paradigm of ProactiveVA.

\subsubsection{Observed Issues \& Suggested Improvements} 

Based on post-study interviews and in-session observations, we identified several usability and interaction issues encountered by participants.

% \begin{itemize}[leftmargin=*]
    
    % \item 
    \revised{\textbf{Intervene timing.} After reviewing the slider adjustment records, we found that 3 users reduced the thinkTime from 3 seconds to 1 second early on, while another 3 did so during the middle of the exploration. One of them later reverted it back to 3 seconds. Overall, adjustments were made between 0 and 2 times. T2 made the adjustment mid-way, feeling overwhelmed by the system’s information load but not realizing the need for the change. T5 adjusted early on due to difficulties with the system, and T6 later reverted to 3 seconds once they had a clearer exploration strategy and wanted less intervention. In general, users suggested that the agent should dynamically adjust the frequency based on the analysis stage and user background. }
    
    % \item 
    \revised{\textbf{Inference precision.}
    We found that the agent proposed an average of 13.5 suggestions per participant in the data exploration stage, with 8.5 of those being executed. 
    However, this does not mean the remaining parts were misjudgments because users often do not notice them while exploring. 
    From the think-aloud and interview, we found participants explicitly mentioned 6 instances. The reason can be summarized into three types. 
    First, T4 and T6 pointed out that the agent does not adapt to the new task if switching quickly. Second, T4 and T2 said that the agent did not associate previous interaction intent with current intent. Lastly, in complex views, the agent might be influenced by the noisy interactions (T1, T3). These insights point to a need for better long-term context management and sensitivity to user interactions.}
    
    \textbf{Control mechanisms \revised{and explanations.}} While Q8 results indicated that our system provides frequency control, users requested more control options. T4, the only participant dissatisfied with the proactive UI agent in Q3 and Q9, commented, \textit{``The agent correctly identifies my needs, but progresses too quickly for my visualization expertise level. I'd prefer to intervene mid-process to steer its reasoning.''} T4 suggested that for novice users, the agent should maintain stronger logical connections between steps to provide more systematic guidance, complementing demand perception. \revised{Additionally, multiple users appreciated the agent’s speed and effectiveness in analysis, but noted that LLMs sometimes generate comprehensive answers in the note comments that are hard to trace back to their source. It would be better to, e.g., brush text and link it more clearly visual interface (T9).}
% \end{itemize}

\section{Discussion and Conclusion}
\label{sec:discussion}

In this paper, we introduce the ProactiveVA framework, where an LLM-based UI agent collaborates with human analysts, infers user intent and provides context-aware assistance autonomously. Throughout the implementation and evaluation, we identified several design trade-offs and directions for future research. We hope that these insights will inspire future work in this field.

\subsection{Design Trade-off}

\textbf{Direct execution vs. Proactive clarification.} 
Our user studies revealed a design tradeoff between direct execution and proactively clarifying with the user~\cite{OhBetter2024a}. When the model is strong and has high confidence for a specific task, direct execution can save the user time and yield good results. However, if the model's capabilities are insufficient or there are misunderstandings, direct execution may lead to errors and waste the user’s time. In such cases, proactively clarifying the user’s intent is a better choice. Future designs could measure the task complexity and the model's capabilities (e.g., confidence and accuracy score)~\cite{han2024towards} to surface the uncertainty for agency allocation.

\textbf{Lightweight tips vs. In-depth guidance.}
One key design tradeoff lies in determining the degree of system guidance~\cite{ceneda2016characterizing, SperrleLotse}. While the current system provides lightweight tips for onboarding, more complex tasks may benefit from in-depth guidance~\cite{perez2025coupling}. However, excessive guidance can overwhelm users if not managed properly. Similarly, when dealing with many notes, users may need to retrace multiple steps to verify details, which increases cognitive load. To address this, dynamic adjustments based on task complexity can help tailor the support. Rather than relying on a single type, developer-defined guidance, a malleable, generative interface could be another way~\cite{Malleable} where guidance is created dynamically through user interaction.

\subsection{Limitations and Future Directions}

\revised{\textbf{Limitation of ProactiveVA.}
While our framework performs well in current VA tasks, its generalizability to other complex systems remains limited. The framework may struggle with tasks involving large-scale datasets or complex task dependencies, requiring further adaptation for specific systems. To enhance the framework's versatility, we could further make our tool into a plug-and-play assistant, which would require VA systems to provide APIs for data transmission and interaction instructions. LLM-based evaluations, while efficient, may misinterpret user intent in unfamiliar or complex scenarios, potentially leading to inaccurate assessments. Additionally, user studies could be improved by focusing on the comparison between user intent and system suggestions, offering a more detailed evaluation of the system's performance. Future research should explore more precise user intent matching and expand evaluation diversity for more robust validation.}

\textbf{\revised{Improving intent inference precision.}}
% \revised{The understanding of long sequence in VA systems still requires in-depth research.}
To address current limitations in intent perception, we identify two key directions for improvement:
Long-term context understanding: While effective for short-term needs, the agent struggles with long-term contexts~\cite{WuMLDT2024}. We suggest integrating summarization methods~\cite{GotzCharacterizing2009}, e.g., through instruction finetuning~\cite{ZhangSummAct2025}, enabling the agent to summarize key milestones and associate tasks timely. \revised{Spatial awareness: User actions in close proximity may confuse the agent’s perception. To enhance spatial awareness, we can leverage multimodal large models~\cite{KilDualView2024} that jointly process HTML structure, interface screenshots, and visualization knowledge. This integration helps the model better understand graphical layout semantics and reduces susceptibility to noisy actions.}

\revised{\textbf{Adaptive support for dynamic user cognition.}
Our study shows that user cognition evolves throughout the analysis process, requiring more adaptive support~\cite{ TankelevitchMetacognitive2024}. For instance, when the task difficulty exceeds their skill level (e.g, on the early exploration stage), they typically require more guidance. As users improve their abilities and gain a better understanding of the agent’s capabilities, they may transition into roles where they collaborate with the agent or guide it~\cite{amershi2014power}. In such cases, the model may reduce the frequency of assistance, focusing on more specific needs. Understanding how users form their mental models and how the agent should adapt to these cognitive changes requires further research. User modeling methods~\cite{yanez2025state} may offer insights to support such adaptivity, both explicit (e.g., onboarding questionnaires~\cite{steichen2013user, carenini2014highlighting}, feedback to suggestions~\cite{LuProactive2024}) and implicit (e.g., interaction pace~\cite{lalle2016predicting, conati2020comparing}).}

\textbf{More fine-grained control of proactive agent.}
Our user studies revealed that while participants valued the ability to adjust the frequency of recommendations and the categories of guidance, they also desired deeper control over the agent's cognitive processes. Unlike conventional reactive systems, proactive UI agents present challenges for real-time human steering. The system lacks finer control mechanisms that would allow users to edit the agent's reasoning mid-process. Thus, a critical research question is how to maintain a human-centered approach while balancing agent autonomy with granular user control~\cite{shneiderman2022human}. \revised{To support this, we can collect more human-in-the-loop feedback\cite{retzlaff2024human}, such as passive cues (e.g., hesitation or avoidance) and demonstrations, or visualize the agent's thinking process~\cite{SubramonyamTexSketch2020, WuPromptChainer2022a}, and design multi-modal controls incorporating interactions~\cite{LinInkSight2023}, texts, and speech~\cite{RosenbergDrawTalking2024}}.

\end{spacing}

\newpage
\acknowledgments{
This work is supported by the Natural Science Foundation of China (NSFC No.62472099) and Ji Hua Laboratory S\&T Program (X250881UG250).
}

\bibliographystyle{abbrv-doi}

\bibliography{reference}

\begin{thebibliography}{10}

\bibitem{amershi2014power}
S.~Amershi, M.~Cakmak, W.~B. Knox, and T.~Kulesza.
\newblock {Power to the People}: The role of humans in interactive machine learning.
\newblock {\em AI magazine}, 35(4):105--120, 2014.

\bibitem{AndrienkoSeeking2022}
N.~Andrienko, G.~Andrienko, S.~Chen, and B.~Fisher.
\newblock Seeking patterns of visual pattern discovery for knowledge building.
\newblock {\em Computer Graphics Forum}, 41(6):124--148, 2022.

\bibitem{BattleCharacterizing2019}
L.~Battle and J.~Heer.
\newblock Characterizing exploratory visual analysis: A literature review and evaluation of analytic provenance in tableau.
\newblock {\em Computer Graphics Forum}, 38(3):145--159, June 2019.

\bibitem{BoukhelifaCase2021}
N.~Boukhelifa, E.~Lutton, and A.~Bezerianos.
\newblock A case study of using analytic provenance to reconstruct user trust in a guided visual analytics system.
\newblock In {\em 2021 IEEE Workshop on TRust and EXpertise in Visual Analytics}, TREX'21, pp. 45--51. IEEE, Oct. 2021.

\bibitem{BrehmerMultiLevel2013}
M.~Brehmer and T.~Munzner.
\newblock A multi-level typology of abstract visualization tasks.
\newblock {\em IEEE Transactions on Visualization and Computer Graphics}, 19(12):2376--2385, Dec. 2013.

\bibitem{BrownFinding2014}
E.~T. Brown, A.~Ottley, H.~Zhao, Q.~Lin, R.~Souvenir, A.~Endert, and R.~Chang.
\newblock Finding waldo: Learning about users from their interactions.
\newblock {\em IEEE Transactions on Visualization and Computer Graphics}, 20(12):1663--1672, Dec. 2014.

\bibitem{carenini2014highlighting}
G.~Carenini, C.~Conati, E.~Hoque, B.~Steichen, D.~Toker, and J.~Enns.
\newblock Highlighting interventions and user differences: Informing adaptive information visualization support.
\newblock In {\em In proceedings of the 2014 CHI Conference on Human Factors in Computing Systems}, CHI '14, pp. 1835--1844, 2014.

\bibitem{ceneda2016characterizing}
D.~Ceneda, T.~Gschwandtner, T.~May, S.~Miksch, H.-J. Schulz, M.~Streit, and C.~Tominski.
\newblock Characterizing guidance in visual analytics.
\newblock {\em IEEE transactions on visualization and computer graphics}, 23(1):111--120, 2016.

\bibitem{ChenNeed2024}
V.~Chen, A.~Zhu, S.~Zhao, H.~Mozannar, D.~Sontag, and A.~Talwalkar.
\newblock Need help? designing proactive ai assistants for programming.
\newblock {\em arXiv:2410.04596}, Oct. 2024.

\bibitem{conati2020comparing}
C.~Conati, S.~Lall{\'e}, M.~A. Rahman, and D.~Toker.
\newblock Comparing and combining interaction data and eye-tracking data for the real-time prediction of user cognitive abilities in visualization tasks.
\newblock {\em ACM Transactions on Interactive Intelligent Systems (TiiS)}, 10(2):1--41, 2020.

\bibitem{CookMixedinitiative2015}
K.~Cook, N.~Cramer, D.~Israel, M.~Wolverton, J.~Bruce, R.~Burtner, and A.~Endert.
\newblock Mixed-initiative visual analytics using task-driven recommendations.
\newblock In {\em 2015 IEEE Conference on Visual Analytics Science and Technology (VAST)}, VAST '15, pp. 9--16. IEEE, Oct. 2015.

\bibitem{CuiDataSite2018a}
Z.~Cui, S.~K. Badam, A.~Yal{\c c}in, and N.~Elmqvist.
\newblock {DataSite}: Proactive visual data exploration with computation of insight-based recommendations.
\newblock {\em arXiv:1802.08621}, Sept. 2018.

\bibitem{DabekGrammarbased2017}
F.~Dabek and J.~J. Caban.
\newblock A grammar-based approach for modeling user interactions and generating suggestions during the data exploration process.
\newblock {\em IEEE Transactions on Visualization and Computer Graphics}, 23(1):41--50, Jan. 2017.

\bibitem{DengSurvey2023a}
Y.~Deng, W.~Lei, W.~Lam, and T.-S. Chua.
\newblock A survey on proactive dialogue systems: Problems, methods, and prospects.
\newblock {\em arXiv:2305.02750}, May 2023.

\bibitem{DengHumancentered2024a}
Y.~Deng, L.~Liao, Z.~Zheng, G.~H. Yang, and T.-S. Chua.
\newblock Towards human-centered proactive conversational agents.
\newblock In {\em Proceedings of the 47th International ACM SIGIR Conference on Research and Development in Information Retrieval}, SIGIR'21, pp. 807--818. ACM, July 2024.

\bibitem{DibiaLIDA2023}
V.~Dibia.
\newblock {LIDA}: A tool for automatic generation of grammar-agnostic visualizations and infographics using large language models.
\newblock {\em arXiv:2303.02927}, Mar. 2023.

\bibitem{GaoFineTuned2024}
L.~Gao, J.~Lu, Z.~Shao, Z.~Lin, S.~Yue, C.~Ieong, Y.~Sun, R.~J. Zauner, Z.~Wei, and S.~Chen.
\newblock Fine-tuned large language model for visualization system: A study on self-regulated learning in education.
\newblock {\em arXiv:2407.20570}, July 2024.

\bibitem{GotzBehaviordriven2009}
D.~Gotz and Z.~Wen.
\newblock Behavior-driven visualization recommendation.
\newblock In {\em Proceedings of the 14th International Conference on Intelligent User Interfaces}, pp. 315--324. ACM, Feb. 2009.

\bibitem{GotzCharacterizing2009}
D.~Gotz and M.~X. Zhou.
\newblock Characterizing users' visual analytic activity for insight provenance.
\newblock {\em Information Visualization}, 8(1):42--55, Jan. 2009.

\bibitem{grant2008dynamics}
A.~M. Grant and S.~J. Ashford.
\newblock The dynamics of proactivity at work.
\newblock {\em Research in organizational behavior}, 28:3--34, 2008.

\bibitem{GuoCase2016}
H.~Guo, S.~R. Gomez, C.~Ziemkiewicz, and D.~H. Laidlaw.
\newblock A case study using visualization interaction logs and insight metrics to understand how analysts arrive at insights.
\newblock {\em IEEE Transactions on Visualization and Computer Graphics}, 22(1):51--60, Jan. 2016.

\bibitem{GuoProactive2011b}
P.~J. Guo, S.~Kandel, J.~M. Hellerstein, and J.~Heer.
\newblock {Proactive Wrangling}: Mixed-initiative end-user programming of data transformation scripts.
\newblock In {\em Proceedings of the 24th Annual ACM Symposium on User Interface Software and Technology}, UIST '24, pp. 65--74. ACM, Oct. 2011.

\bibitem{Superstore_embedded}
D.~Hagen.
\newblock Superstore dashboard - profitability overview.
\newblock \url{https://public.tableau.com/app/profile/dave.hagen/viz/Superstore_embedded_800x800/Overview}, Jan 2023.
\newblock Accessed: Jul. 2, 2025.

\bibitem{han2024towards}
J.~Han, W.~Buntine, and E.~Shareghi.
\newblock Towards uncertainty-aware language agent.
\newblock {\em arXiv preprint arXiv:2401.14016}, 2024.

\bibitem{HarrisonJupyterLab2024}
G.~Harrison, K.~Bryson, A.~E.~B. Bamba, L.~Dovichi, A.~H. Binion, A.~Borem, and B.~Ur.
\newblock Jupyterlab in retrograde: Contextual notifications that highlight fairness and bias issues for data scientists.
\newblock In {\em Proceedings of the CHI Conference on Human Factors in Computing Systems}, CHI '24, pp. 1--19. ACM, May 2024.

\bibitem{HeerInteractive2012}
J.~Heer and B.~Shneiderman.
\newblock Interactive dynamics for visual analysis.
\newblock {\em Commun. ACM}, 55(4):45–54,  10 pages, Apr. 2012.

\bibitem{HorvitzPrinciples1999a}
E.~Horvitz.
\newblock Principles of mixed-initiative user interfaces.
\newblock In {\em In proceedings of the 1999 CHI Conference on Human Factors in Computing Systems}, CHI '99, pp. 159--166. ACM Press, 1999.

\bibitem{kadoma2024role}
K.~Kadoma, M.~Aubin Le~Quere, X.~J. Fu, C.~Munsch, D.~Metaxa, and M.~Naaman.
\newblock The role of inclusion, control, and ownership in workplace ai-mediated communication.
\newblock In {\em Proceedings of the CHI Conference on Human Factors in Computing Systems}, CHI '24, pp. 1--10, 2024.

\bibitem{KeimVisual2008c}
D.~A. Keim, F.~Mansmann, J.~Schneidewind, J.~Thomas, and H.~Ziegler.
\newblock {Visual Analytics}: Scope and challenges.
\newblock In S.~J. Simoff, M.~H. B{\"o}hlen, and A.~Mazeika, eds., {\em Visual Data Mining}, vol. 4404, pp. 76--90. Springer Berlin Heidelberg, 2008.

\bibitem{KilDualView2024}
J.~Kil, C.~H. Song, B.~Zheng, X.~Deng, Y.~Su, and W.-L. Chao.
\newblock Dual-view visual contextualization for web navigation.
\newblock {\em arXiv:2402.04476}, Mar. 2024.

\bibitem{KimPhenoFlow2024c}
J.~Kim, S.~Lee, H.~Jeon, K.-J. Lee, H.-J. Bae, B.~Kim, and J.~Seo.
\newblock {PhenoFlow}: A human-llm driven visual analytics system for exploring large and complex stroke datasets.
\newblock {\em IEEE Transactions on Visualization and Computer Graphics}, pp. 1--11, 2024.

\bibitem{LaiAutoWebGLM2024}
H.~Lai, X.~Liu, I.~L. Iong, S.~Yao, Y.~Chen, P.~Shen, H.~Yu, H.~Zhang, X.~Zhang, Y.~Dong, and J.~Tang.
\newblock {AutoWebGLM}: A large language model-based web navigating agent.
\newblock {\em arXiv:2404.03648}, Oct. 2024.

\bibitem{lalle2016predicting}
S.~Lall{\'e}, C.~Conati, and G.~Carenini.
\newblock Predicting confusion in information visualization from eye tracking and interaction data.
\newblock In {\em IJCAI}, pp. 2529--2535, 2016.

\bibitem{LiDiverse2022}
Y.~Li, Y.~Qi, Y.~Shi, Q.~Chen, N.~Cao, and S.~Chen.
\newblock Diverse interaction recommendation for public users exploring multi-view visualization using deep learning.
\newblock {\em IEEE Transactions on Visualization and Computer Graphics}, pp. 1--11, 2022.

\bibitem{LinInkSight2023}
Y.~Lin, H.~Li, L.~Yang, A.~Wu, and H.~Qu.
\newblock {InkSight}: Leveraging sketch interaction for documenting chart findings in computational notebooks.
\newblock {\em IEEE Transactions on Visualization and Computer Graphics}, 30(1):944--954, 2023.

\bibitem{liu2024deepseek}
A.~Liu, B.~Feng, B.~Xue, B.~Wang, B.~Wu, C.~Lu, C.~Zhao, C.~Deng, C.~Zhang, C.~Ruan, et~al.
\newblock Deepseek-v3 technical report.
\newblock {\em arXiv preprint arXiv:2412.19437}, 2024.

\bibitem{liu2024ava}
S.~Liu, H.~Miao, Z.~Li, M.~Olson, V.~Pascucci, and P.-T. Bremer.
\newblock {AVA}: Towards autonomous visualization agents through visual perception-driven decision-making.
\newblock In {\em Computer Graphics Forum}, vol.~43, p. e15093. Wiley Online Library, 2024.

\bibitem{LiuComPeer2024}
T.~Liu, H.~Zhao, Y.~Liu, X.~Wang, and Z.~Peng.
\newblock {ComPeer}: A generative conversational agent for proactive peer support.
\newblock {\em UIST '24: Proceedings of the 37th Annual ACM Symposium on User Interface Software and Technology}, pp. 1--22, 2024.

\bibitem{LiuProactive2024}
X.~B. Liu, S.~Fang, W.~Shi, C.-S. Wu, T.~Igarashi, and X.~A. Chen.
\newblock Proactive conversational agents with inner thoughts.
\newblock {\em arXiv:2501.00383}, Dec. 2024.

\bibitem{LuProactive2024}
Y.~Lu, S.~Yang, C.~Qian, G.~Chen, Q.~Luo, Y.~Wu, H.~Wang, X.~Cong, Z.~Zhang, Y.~Lin, W.~Liu, Y.~Wang, Z.~Liu, F.~Liu, and M.~Sun.
\newblock {Proactive Agent}: Shifting llm agents from reactive responses to active assistance.
\newblock {\em arXiv:2410.12361}, Dec. 2024.

\bibitem{Malleable}
B.~Min, A.~Chen, Y.~Cao, and H.~Xia.
\newblock Malleable overview-detail interfaces.
\newblock In {\em Proceedings of the 2025 CHI Conference on Human Factors in Computing Systems}, CHI '25,  article no. 688,  25 pages. Association for Computing Machinery, 2025.

\bibitem{musleh2025trustme}
M.~Musleh, R.~G. Raidou, and D.~Ceneda.
\newblock {TrustME}: A context-aware explainability model to promote user trust in guidance.
\newblock {\em IEEE Transactions on Visualization and Computer Graphics}, 2025.

\bibitem{OhBetter2024a}
J.~Oh, W.~Kim, S.~Kim, H.~Im, and S.~Lee.
\newblock {Better to Ask Than Assume}: Proactive voice assistants' communication strategies that respect user agency in a smart home environment.
\newblock In {\em Proceedings of the CHI Conference on Human Factors in Computing Systems}, CHI '24, pp. 1--17. ACM, May 2024.

\bibitem{Peng2021Mixed}
L.~Peng, Y.~Zhao, Y.~Hou, Q.~Wang, S.~Shen, X.~Lai, J.~Gao, J.~Dong, Z.~Lin, and S.~Chen.
\newblock Mixed-initiative visual exploration of social media text and events.
\newblock In {\em Proceedings of the IEEE Conference on Visualization and Visual Analytics}, VIS '21, 2021.

\bibitem{peng2019design}
Z.~Peng, Y.~Kwon, J.~Lu, Z.~Wu, and X.~Ma.
\newblock Design and evaluation of service robot's proactivity in decision-making support process.
\newblock In {\em proceedings of the 2019 CHI conference on human factors in computing systems}, CHI '19, pp. 1--13, 2019.

\bibitem{perez2025coupling}
I.~P{\'e}rez-Messina, M.~Angelini, D.~Ceneda, C.~Tominski, and S.~Miksch.
\newblock Coupling guidance and progressiveness in visual analytics.
\newblock In {\em Computer Graphics Forum}, p. e70115. Wiley Online Library, 2025.

\bibitem{perez2023guided}
I.~P{\'e}rez-Messina, D.~Ceneda, and S.~Miksch.
\newblock Guided visual analytics for image selection in time and space.
\newblock {\em IEEE Transactions on Visualization and Computer Graphics}, 30(1):66--75, 2023.

\bibitem{PisterIntegrating2021a}
A.~Pister, P.~Buono, J.-D. Fekete, C.~Plaisant, and P.~Valdivia.
\newblock Integrating prior knowledge in mixed-initiative social network clustering.
\newblock {\em IEEE Transactions on Visualization and Computer Graphics}, 27(2):1775--1785, Feb. 2021.

\bibitem{retzlaff2024human}
C.~O. Retzlaff, S.~Das, C.~Wayllace, P.~Mousavi, M.~Afshari, T.~Yang, A.~Saranti, A.~Angerschmid, M.~E. Taylor, and A.~Holzinger.
\newblock Human-in-the-loop reinforcement learning: A survey and position on requirements, challenges, and opportunities.
\newblock {\em Journal of Artificial Intelligence Research}, 79:359--415, 2024.

\bibitem{RosenbergDrawTalking2024}
K.~T. Rosenberg, R.~H. Kazi, L.-Y. Wei, H.~Xia, and K.~Perlin.
\newblock {DrawTalking}: Building interactive worlds by sketching and speaking.
\newblock {\em arXiv:2401.05631}, Aug. 2024.

\bibitem{ross2023programmer}
S.~I. Ross, F.~Martinez, S.~Houde, M.~Muller, and J.~D. Weisz.
\newblock The programmer’s assistant: Conversational interaction with a large language model for software development.
\newblock In {\em Proceedings of the 28th International Conference on Intelligent User Interfaces}, pp. 491--514, 2023.

\bibitem{ShinDrillboards2024}
S.~Shin, I.~Na, and N.~Elmqvist.
\newblock Drillboards: Adaptive visualization dashboards for dynamic personalization of visualization experiences.
\newblock {\em arXiv:2410.12744}, Oct. 2024.

\bibitem{shneiderman2022human}
B.~Shneiderman.
\newblock {\em {Human-centered AI}}.
\newblock Oxford University Press, 2022.

\bibitem{SperrleLotse}
F.~Sperrle, D.~Ceneda, and M.~{El-Assady}.
\newblock {Lotse}: A practical framework for guidance in visual analytics.
\newblock {\em IEEE Transactions on Visualization and Computer Graphics}, 29(1):1124--1134, 2023.

\bibitem{SperrleLearning2021}
F.~Sperrle, H.~Sch{\"a}fer, D.~Keim, and M.~{El-Assady}.
\newblock Learning contextualized user preferences for co-adaptive guidance in mixed-initiative topic model refinement.
\newblock {\em Computer Graphics Forum}, 40(3):215--226, 2021.

\bibitem{SrinivasanDashboard2024}
A.~Srinivasan, J.~Purich, M.~Correll, L.~Battle, V.~Setlur, and A.~Crisan.
\newblock From dashboard zoo to census: A case study with tableau public.
\newblock {\em IEEE Transactions on Visualization and Computer Graphics}, pp. 1--15, 2024.

\bibitem{steichen2013user}
B.~Steichen, G.~Carenini, and C.~Conati.
\newblock User-adaptive information visualization: using eye gaze data to infer visualization tasks and user cognitive abilities.
\newblock In {\em Proceedings of the 2013 International Conference on Intelligent User Interfaces}, IUI '23, pp. 317--328, 2013.

\bibitem{StoiberPerspectives2022}
C.~Stoiber, D.~Ceneda, M.~Wagner, V.~Schetinger, T.~Gschwandtner, M.~Streit, S.~Miksch, and W.~Aigner.
\newblock Perspectives of visualization onboarding and guidance in va.
\newblock {\em Visual Informatics}, 6(1):68--83, Mar. 2022.

\bibitem{StolperProgressive2014}
C.~D. Stolper, A.~Perer, and D.~Gotz.
\newblock Progressive visual analytics: User-driven visual exploration of in-progress analytics.
\newblock {\em IEEE Transactions on Visualization and Computer Graphics}, 20(12):1653--1662, Dec. 2014.

\bibitem{SubramonyamTexSketch2020}
H.~Subramonyam, C.~Seifert, P.~Shah, and E.~Adar.
\newblock {texSketch}: Active diagramming through pen-and-ink annotations.
\newblock In {\em Proceedings of the 2020 CHI Conference on Human Factors in Computing Systems}, CHI '20, pp. 1--13. ACM, Apr. 2020.

\bibitem{TabalbaArticulatePro2024}
R.~Tabalba, C.~J. Lee, G.~Tran, N.~Kirshenbaum, and J.~Leigh.
\newblock {ArticulatePro}: A comparative study on a proactive and non-proactive assistant in a climate data exploration task.
\newblock {\em arXiv:2409.10797}, Sept. 2024.

\bibitem{Tableau2003Embedding}
Tableau.
\newblock Tableau embedding api.
\newblock \url{https://help.tableau.com/current/api/embedding_api/en-us/index.html}, 2003.
\newblock Accessed: Jul. 2, 2025.

\bibitem{TankelevitchMetacognitive2024}
L.~Tankelevitch, V.~Kewenig, A.~Simkute, A.~E. Scott, A.~Sarkar, A.~Sellen, and S.~Rintel.
\newblock The metacognitive demands and opportunities of generative ai.
\newblock In {\em Proceedings of the CHI Conference on Human Factors in Computing Systems}, CHI '24, pp. 1--24. ACM.

\bibitem{thomas2005illuminating}
J.~Thomas.
\newblock {\em {Illuminating the Path}: The Research and Development Agenda for Visual Analytics}.
\newblock IEEE, 2005.

\bibitem{TianChartGPT2023b}
Y.~Tian, W.~Cui, D.~Deng, X.~Yi, Y.~Yang, H.~Zhang, and Y.~Wu.
\newblock {ChartGPT}: Leveraging llms to generate charts from abstract natural language.
\newblock {\em arXiv:2311.01920}, Nov. 2023.

\bibitem{wall2019markov}
E.~Wall, A.~Arcalgud, K.~Gupta, and A.~Jo.
\newblock A markov model of users’ interactive behavior in scatterplots.
\newblock In {\em 2019 IEEE visualization conference}, pp. 81--85. IEEE, 2019.

\bibitem{weng2025insightlens}
L.~Weng, X.~Wang, J.~Lu, Y.~Feng, Y.~Liu, H.~Feng, D.~Huang, and W.~Chen.
\newblock {InsightLens}: Augmenting llm-powered data analysis with interactive insight management and navigation.
\newblock {\em IEEE Transactions on Visualization and Computer Graphics}, 2025.

\bibitem{WuPromptChainer2022a}
T.~Wu, E.~Jiang, A.~Donsbach, J.~Gray, A.~Molina, M.~Terry, and C.~J. Cai.
\newblock {PromptChainer}: Chaining large language model prompts through visual programming.
\newblock In {\em Extended Abstracts of the 2022 CHI Conference on Human Factors in Computing Systems}, CHI EA '22,  article no. 359,  10 pages, pp. 1--10. Association for Computing Machinery, 2022.

\bibitem{WuMLDT2024}
Y.~Wu, J.~Zhang, N.~Hu, L.~Tang, G.~Qi, J.~Shao, J.~Ren, and W.~Song.
\newblock {MLDT}: Multi-level decomposition for complex long-horizon robotic task planning with open-source large language model.
\newblock {\em arXiv:2403.18760}, Apr. 2024.

\bibitem{XuSurvey2020}
K.~Xu, A.~Ottley, C.~Walchshofer, M.~Streit, R.~Chang, and J.~Wenskovitch.
\newblock Survey on the analysis of user interactions and visualization provenance.
\newblock {\em Computer Graphics Forum}, 39(3):757--783, June 2020.

\bibitem{yanez2025state}
F.~Yanez, C.~Conati, A.~Ottley, and C.~Nobre.
\newblock The state of the art in user-adaptive visualizations.
\newblock In {\em Computer Graphics Forum}, vol.~44, p. e15271. Wiley Online Library, 2025.

\bibitem{YaoReAct2023}
S.~Yao, J.~Zhao, D.~Yu, N.~Du, I.~Shafran, K.~Narasimhan, and Y.~Cao.
\newblock {ReAct}: Synergizing reasoning and acting in language models.
\newblock {\em arXiv:2210.03629}, Mar. 2023.

\bibitem{ZarghamUnderstanding2022a}
N.~Zargham, L.~Reicherts, M.~Bonfert, S.~T. Voelkel, J.~Schoening, R.~Malaka, and Y.~Rogers.
\newblock Understanding circumstances for desirable proactive behaviour of voice assistants: The proactivity dilemma.
\newblock In {\em Proceedings of the 4th Conference on Conversational User Interfaces}, pp. 1--14. ACM, July 2022.

\bibitem{zhang2024ufo}
C.~Zhang, L.~Li, S.~He, X.~Zhang, B.~Qiao, S.~Qin, M.~Ma, Y.~Kang, Q.~Lin, S.~Rajmohan, et~al.
\newblock {UFO}: A ui-focused agent for windows os interaction.
\newblock {\em arXiv preprint arXiv:2402.07939}, 2024.

\bibitem{ZhangSummAct2025}
G.~Zhang, M.~A.~N. Ahmed, Z.~Hu, and A.~Bulling.
\newblock {SummAct}: Uncovering user intentions through interactive behaviour summarisation.
\newblock In {\em Proceedings of the 2025 CHI Conference on Human Factors in Computing Systems}, CHI '25, pp. 1--17. ACM, Apr. 2025.

\bibitem{ZhangAskbeforePlan2024}
X.~Zhang, Y.~Deng, Z.~Ren, S.-K. Ng, and T.-S. Chua.
\newblock Ask-before-plan: Proactive language agents for real-world planning.
\newblock {\em arXiv:2406.12639}, June 2024.

\bibitem{ZhaoLightVA2024}
Y.~Zhao, J.~Wang, L.~Xiang, X.~Zhang, Z.~Guo, C.~Turkay, Y.~Zhang, and S.~Chen.
\newblock {LightVA}: Lightweight visual analytics with llm agent-based task planning and execution.
\newblock {\em IEEE Transactions on Visualization and Computer Graphics}, pp. 1--13, 2024.

\bibitem{ZhaoLEVA2024}
Y.~Zhao, Y.~Zhang, Y.~Zhang, X.~Zhao, J.~Wang, Z.~Shao, C.~Turkay, and S.~Chen.
\newblock {LEVA}: Using large language models to enhance visual analytics.
\newblock {\em IEEE Transactions on Visualization and Computer Graphics}, pp. 1--17, 2024.

\bibitem{zhao2024generating}
Z.~Zhao and Z.~Dou.
\newblock Generating multi-turn clarification for web information seeking.
\newblock In {\em Proceedings of the ACM on Web Conference 2024}, WWW '24, pp. 1539--1548, 2024.

\bibitem{ZhouWebArena2024}
S.~Zhou, F.~F. Xu, H.~Zhu, X.~Zhou, R.~Lo, A.~Sridhar, X.~Cheng, T.~Ou, Y.~Bisk, D.~Fried, U.~Alon, and G.~Neubig.
\newblock {WebArena}: A realistic web environment for building autonomous agents.
\newblock {\em arXiv:2307.13854}, Apr. 2024.

\end{thebibliography}

\end{document}